\documentclass[twocolumn]{aa} 
\usepackage{graphicx}
\usepackage{txfonts}
\begin{document}
\title{Recovering the star formation rate in the solar neighborhood}

   \subtitle{} 

   \author{M. Cignoni\inst{1,3}, S. Degl'Innocenti\inst{1,2}, 
P. G. Prada Moroni\inst{1,2}, S. N. Shore\inst{1,2}}

 \offprints{M. Cignoni}

   \institute{
(1) Dipartimento di Fisica ``Enrico Fermi'', Universit\`a di Pisa, largo Pontecorvo 3, Pisa I-56127 Italy;\\
(2) INFN - Sezione di Pisa, largo Pontecorvo 3, Pisa I-56127,
     Italy;\\
(3) Osservatorio Astronomico Di Capodimonte, Via Moiariello 16,
     80131 Napoli, Italy}
    \titlerunning{The SFR in the solar neighborhood}
   \date{Received; accepted }

   \abstract{
{This paper develops a method for obtaining the star formation
     histories of a mixed, resolved population through the use of
     color-magnitude diagrams (CMDs).  The method provides insight into the
     local star formation rate, analyzing the observations of the Hipparcos
     satellite through a comparison with synthetic CMDs computed for different
     histories with an updated stellar evolution library.}  
{Parallax and
     photometric uncertainties are included explicitly and corrected using the
     Bayesian Richardson-Lucy algorithm.  We first describe our verification
     studies using artificial data sets.  From this sensitivity study, the
     critical factors determining the success of a recovery for a known star
     formation rate are a partial knowledge of the IMF and the age-metallicity
     relation, and sample contamination by clusters and moving groups (special
     populations whose histories are different than that of the whole sample).
     Unresolved binaries are less important impediments.  We highlight how
     these limit the method.}
  {For the real field sample, complete to $M_V <
     3.5$, we find that the solar neighborhood star formation rate has a
     characteristic timescale for variation of about 6 Gyr, with a maximum
     activity close to 3 Gyr ago. The similarity of this finding with column
     integrated star formation rates may indicate a global origin, possibly a
     collision with a satellite galaxy.  We also discuss applications of this
     technique to general photometric surveys of other complex systems
     (e.g. Local Group dwarf galaxies) where the distances are well known.}
     } 
   \maketitle

%

\section{Introduction}
Since the pioneering works of Bahcall \& Soneira (\cite{bah}) and Gilmore \&
Reid (\cite{gil83}), the comparison with deep star counts has been a primary
method to disentangle the stellar components of the Milky Way. However, the
number of parameters and the observational uncertainties involved in this kind
of analysis are often prohibitive.  In contrast, nearby stars are a well
defined sample of disk stars and the distances are now well known so one can
extract detailed information about the local star formation rate and how this
is connected with the whole Galactic disk.

In this framework, the Hipparcos catalog provides a unique opportunity. Before
Hipparcos, the local stellar population of bright stars, in main sequence or
in red giant phase, was poorly represented; moreover, the distance
uncertainties washed out most of the fine structure of the color-magnitude
diagrams (CMDs).  After Hipparcos, it was possible to study the
color-magnitude diagram for local stars in a statistical sense, beyond the
simple comparison between evolutionary tracks and single stars, a method for
which is the subject of this paper.

Published studies based on the Hipparcos local sample have concentrated on two
approaches. One is the direct comparison between data and artificial CMDs
using a likelihood function, e.g.  Bertelli \& Nasi (\cite{bert}) and
Schr\"oder \& Pagel (\cite{scho}). The second is the Bayesian approach,
e.g. by Hernandez et al. (\cite{her2}) and Vergely et al. (\cite{verg}).

Here we suggest an hybrid technique (for an over-complete description see 
Cignoni 2006). First, we adopt a {Bayesian treatment of the observational
  uncertainties}: the CMD is converted into an image (binning process) and a
  Richardson-Lucy algorithm is used to clean the data (see Cignoni \& Shore
  2006 for details). This ``cleaned'' Hipparcos CMD can then be used 
to recover the local star formation rate (SFR). This is done in
  different steps, as described in the following sections: (1) an 
ensemble of synthetic CMDs is generated using 
Monte Carlo simulations; (2) a likelihood function is minimized for the comparison between theory
  and observation; and (3) a confidence limit of the result is evaluated with a bootstrap technique.

In section 2 we describe the properties
of the selected volume complete sample and how we removed the known
stellar clusters, moving groups, and associations. Sections 3 and 4 describe
the method: physical inputs adopted for the stellar evolutionary calculations,
Monte Carlo generations of artificial CMDs, likelihood function and
bootstrap. In section 5 we apply the algorithm to artificial CMDs, showing
which parameters are critical for recovering the star formation rate. 
Section 6 shows the results for the Hipparcos data. Section 7 discusses the
dependence of the recovered SFR from a kinematic selection of the
data. Finally in section 9 the results are discussed and compared with previous works available in the literature.

  \section{Sample selection}
The Hipparcos mission observed objects to a limiting magnitude of about
$V=12.5$ mag, with a completeness limit that depends on Galactic latitude $b$
and spectral type (see e.g. Perryman et al. \cite{per}):
$V<7.9+1.1\sin |{b}|$ for spectral types earlier or equal to G5, $V<7.3
+1.1\sin |{b}|$ for spectral types later than G5.

\begin{figure} 
\centering \includegraphics[width=7cm]{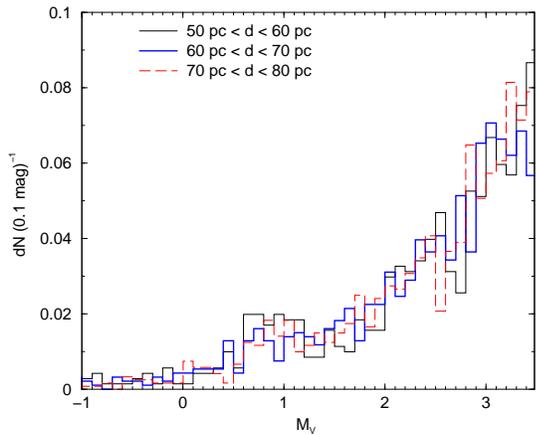} 
 \caption{Luminosity functions (in absolute visual magnitude) for stars
 selected in distance (as labeled). All the stars are brighter than $V=8$.}
\label{malmb}
\end{figure}

\begin{figure}[]
\centering \includegraphics[width=7cm]{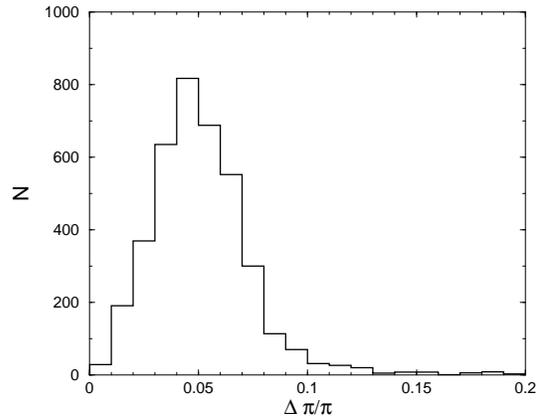}  
\label{precision}
\caption{Distribution of the parallax precision ($\Delta \pi/\pi$) for Hipparcos stars within 80
  pc and brighter than $V=8$.}
\end{figure}

For a volume complete sample, we chose stars within 80 pc of the Sun and brighter than
$V=8$ in visual apparent magnitude, corresponding to a minimum absolute visual 
magnitude $M_V=3.5$. Considering that the formal completeness
limit is dependent by galactic latitude and spectral type, we checked if the
sample is still complete against the Malmquist bias (in a magnitude limited
sample, the brighter stars are statistically over-represented). The bias was quantified 
by comparing the luminosity functions for
subsamples with different heliocentric distances. Figure \ref{malmb} shows the
luminosity functions for stars with different distances from the Sun at
intervals of 10 pc.  
Using a Kolmogorov-Smirnov test, the hypothesis that the luminosity functions
are realizations of the same distribution (for the range $-1<M_V<3.5$) cannot
be rejected with a probability of 10\%. Thus, the sample selected at $V=8$ should be complete up to 80 pc and
$M_V=3.5$, which is ensured by selecting Hipparcos stars that are brighter
than $V \sim 8$ and within $80$ pc from the Sun. This sample contains about $4000$ objects with
a parallax error generally better than 10\% (see Fig. \ref{precision}).

\subsection{Clusters contamination}
\begin{figure}
\centering \includegraphics[width=7cm]{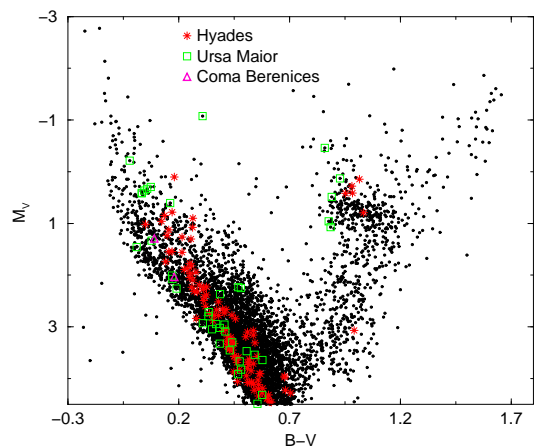} 
 \caption{The selected Hipparcos sample where members of associations are shown.}
\label{asso}
\end{figure}

We are interested in the local star formation activity as it reflects that of the whole 
disk.  In 
contrast, nearby clusters and associations are groups of stars that are 
not dynamically
mixed with the field and which - having been formed together in large numbers
at specific times - do not represent the {\it mean properties} of the solar
neighborhood. 
Figure \ref{asso} shows the identified associations members within 80 pc and
brighter than $V=8$. The most significant contamination is by the Hyades
cluster with about 120 identified members.  

We will hence call ``the Hipparcos sample'' the set defined after that the identified cluster members are eliminated. 

\section{The model}
\subsection{Theoretical inputs}
\label{presc}
To recover the star formation history from the observational CMD
one needs to model a synthetic population with various theoretical
ingredients. Using a Monte Carlo algorithm (CERN library), masses and ages are extracted according assumed 
initial mass functions (IMFs) and star
  formation rates (SFRs). Then, a suitable age-metallicity relation (AMR) is adopted. 
The extracted synthetic stars are placed in the CMD by
  interpolation among the adopted stellar evolution tracks. In order to take
  into account the presence of binary stars, a chosen fraction of these stars
  are assumed as binaries and coupled with a companion star. 

The code relies on a set of evolutionary computations covering
with a fine grid the mass range $\approx\,0.1$ to $\approx\,7 M_{\odot}$, with
metallicities from $Z=0.001$ to $Z=0.03$.
The amount of original helium abundance has been chosen by assuming a
primordial helium abundance $Y_P=0.23$ and $\Delta Y/ \Delta Z \sim 2$ (see
e.g. Pagel \& Portinari \cite{pag}, Castellani, Degl'Innocenti \& Marconi \cite{cast99a}). 
The adopted color transformation are from Castelli, Gratton \& Kurucz (1997)

For masses greater than $0.5\, M_{\odot}$ we use the Cariulo et
al. (\cite{car}), Castellani et al. (\cite{cast03}), Castellani,
Degl'Innocenti \& Marconi (\cite{cast99a}) evolutionary tracks
(partially available at the URL:
http://astro.df.unipi.it/SAA/PEL/Z0.html). The input physics adopted
in the models are described in Cariulo et al. (\cite{car}).
Convective regions,  identified following the Schwarzschild criterion,
are treated with the mixing length formalism in which the mixing
length parameter $\alpha$ defines the ratio between the mixing length
and the local pressure scale height; we have uniformly adopted $\alpha =1.9$, which has been calibrated in a way to reproduce, with the
adopted color transformations, the observed RG branch color of the
galactic globular clusters and young globulars in the LMC (Cariulo et
al. \cite{car}, Castellani et al. \cite{cast03}, Brocato et
al. \cite{broc}). The solar mixture adopted for the calculations is
$[Z/X]_{\odot}=0.0245$ (Grevesse \& Noels \cite{greve}, GN93). We use
throughout the canonical assumption of inefficient overshooting and
the He burning structures are calculated according to the
prescriptions of canonical semiconvection induced by the penetration
of convective elements in the radiative region (Castellani et
al. \cite{cast85}).

Less massive stars ($0.5\, M_{\odot}<\,M\,<\,0.7\, M_{\odot}$), whose
evolutionary times are longer than the Hubble time, have been evolved
up to central H exhaustion. For very low mass stars ($M<0.5\,
M_{\odot}$) we used the Zero Age Main Sequence positions by Baraffe et
al. (\cite{bar7},\cite{bar8}).

\subsection{Artificial CMDs}
\label{stepping}
The so called ``forward'' procedure for obtaining a SFR that can generate an
observed photometric sample diagram is to produce artificial CMDs to be
compared with the observations.  The first technical problem of a similar
approach is the time spent for the Monte Carlo generation of a CMD for each
SFR.  Both the data and the artificial photometry are stored in a color
magnitude grid, each bin of which contains the number of stars observed or
predicted to be in it.  The SFR with the higher probability of generating the
data is chosen by means of a likelihood test.  So, to explore a sufficiently
wide number of star formations, it is necessary to construct a basis set, each
of partial CMDs\footnote{In contrast to the isochrones used for cluster
simulations, these partial CMDs form a statistically {\it fuzzy} set in the
sense that they must span a range of color, luminosity, and abundance
depending on the mass function adopted for the model. This appears more than
an analogy; it may be possible to apply some of the methods already developed
in this field to study even deeper survey fields where the data are less well
constrained than the Hipparcos sample.} with $\approx 10^5$ stars per step
star formation, uniform in a given time interval and zero elsewhere. The step
functions must be exhaustive (the sum covering the whole Hubble time) and they
cannot overlap.

Thus, for each combination of IMF and AMR, the CMD
corresponding to any SFR is computed as a linear
combination of the partial CMDs:

\begin{eqnarray}  
&&m_{i}=\sum_{j}r_{j}\,c_{ij}\label{rl3}
\end{eqnarray} 

where $m_{i}$ is the number of star in the final CMD in bin $i$, $r_{j}$ is the 
star formation rate for the partial CMD $j$, and
$c_{ij}$ is the number of stars in the bin $i$ owning to the partial 
CMD $j$. 

The number of Monte Carlo simulations is reduced to the number of partial
CMDs.  This method has been already applied by several authors (see
e.g. Aparicio, Gallart \& Bertelli \cite{apa1}-\cite{apa2}, Gallart et
al. \cite{gal}, Bertelli \& Nasi \cite{bert}).  The duration of each star
formation interval is chosen depending on the timescale of the typical stellar
population involved, in order to properly sample even those stars with rapid
evolutionary changes. Thus, we have chosen star formation steps of 0.5 Gyr for
stars younger than 2 Gyr, while for the later epoches the duration is
increased (1 Gyr for stars with age between 2 and 4 Gyr, 2 Gyr for older
stars).
\section{The comparison between theory and observation}
\subsection{Choosing the grid}
Choosing the grid dimension for the CMD binning is an essential step.
If the CMD bins are too small, the histogram fluctuations are too large and 
it's more difficult to recover the underlying  SFR.  If the
bin size is too large one can loose information. 

A first rule for choosing the grid uses from the typical masses involved
in the sample: massive stars have shorter
life time (and the corresponding CMD regions are underpopulated) and it is thus suitable to use
a larger bin size in order to map their history. 
Another limit is the evolutionary phase of the star mapping the CMD: 
after the main sequence, the partial CMDs become nearly degenerate and 
consequently the grid needs to be finer.
The adopted binning must arise by numerical simulations for the specific
problem to check the sensitivity of the algorithm to the various choices.

\subsection{Searching for the ``best model''}
To quantify the similarity among CMDs, we chose a
Poisson based likelihood function:
\begin{eqnarray}  
\chi_P= \sum_{i=1}^{N
    bin} n_{i}\ln\frac{n_{i}}{m_{i}}-n_{i}+m_{i}
\label{pois1} 
\end{eqnarray}
where $m_{i}$ and $n_{i}$ are the model and the data histogram in the
i-bin. This ``likelihood'' is considered as a mere ``distance'' to be
minimized, while \emph{the acceptance level of a solution is estimated using a
bootstrap technique}. 

Our model depends on several parameters (10 coefficients of the star formation
rate), thus, the problem is to move
within this multi-dimensional parameter space and to search for the combination of
parameters that minimize $\chi_p$. For this task we implement the
Nelder-Mead simplex method (Nelder \& Mead, \cite{neld}). 
The main problem of the simplex method is the efficency: the presence of
many local minima can prevent reaching a real global minimum for $\chi_P$.
 To improve the efficency we add a
 logarithmically distributed random variable to each vertex of the
 simplex, the minimization algorithm is re-started each time a
 ``global minimum'' is found, and the new departure is randomly chosen (in order to avoid the
dependence by the initial guess) in the parameter space to obtain a class of
best values. The final best parameter is the smallest among the best
values. The restart process is stopped when this ``minimum value'' no longer changes.  

\subsection{Confidence intervals}

The bootstrap method is commonly used to estimate confidence intervals. 
In empirical bootstrap simulations one processes the original data set $N$ times (copies) so each of the original $n$ data points 
is sampled with replacement \footnote{In a random sample with replacement,
  each observation in the data set has an equal chance of being selected and
  it can
  be selected over and over again.} and with equal probability of being sampled.
One finally obtains $N$ different data sets, each with $n$
data points. Because of the replacements, some values in each data set are repeated, 
while others are lacking. This mimics the
observational process: if the observational data is representative of the
underlying distribution, the data produced with replacements are
copies of the original one with local crowding or sparseness.
In practice, the method uses the bootstrapped copies imposing the same
minimization procedure as it would be performed on the real data set. The result
will be a set of ``best'' parameters. 
The confidence interval is then the interval that
contains a defined percentage of this parameter distribution.

\section{Sensitivity tests with artificial data}

Applying the method to artificial data, we have tested:

\begin{enumerate}

\item how the completeness limit, that fixes the boundaries in
magnitude of  the CMD used for the analysis, affects the
result. Different zones in the CMD give information about different
epochs of star formation. The completeness limit determines our ``zone
of ignorance'' for the underlying SFR.

\item how sensitive the recovered SFR is to parametric functions such as the
  IMF, AMR, binary population.  This procedure may highlight parameters whose
  values need to be known very accurately in advance;

\item how the contamination of accidental intruders (e.g. stellar
cluster stars) in the sample, could lead to a biased SFR.

\end{enumerate}
For all tests we have fixed the grid bin size. Young
partial CMDs comprise massive stars so the spanned region is broad (the more
massive the star is, the longer its excursion in color to reach the red giant
branch) and poorly populated: small bins contain few stars and the recovered SFR
suffers of low number statistics. Old partial CMDs are
composed by less massive stars so a narrower bin size is required to
distinguish different SFRs. After some trials we found that a bin size of 0.05 mag, both for
the color and absolute magnitude, is the best compromise.
{\it Hereafter, we call ``artificial data'' each synthetic CMD mimicking the
Hipparcos data while the theoretical CMD is simply called the ``model''}.

\subsection{Completeness limits}
\begin{figure*}
\centering
\includegraphics[width=6.5cm,height=5.5cm]{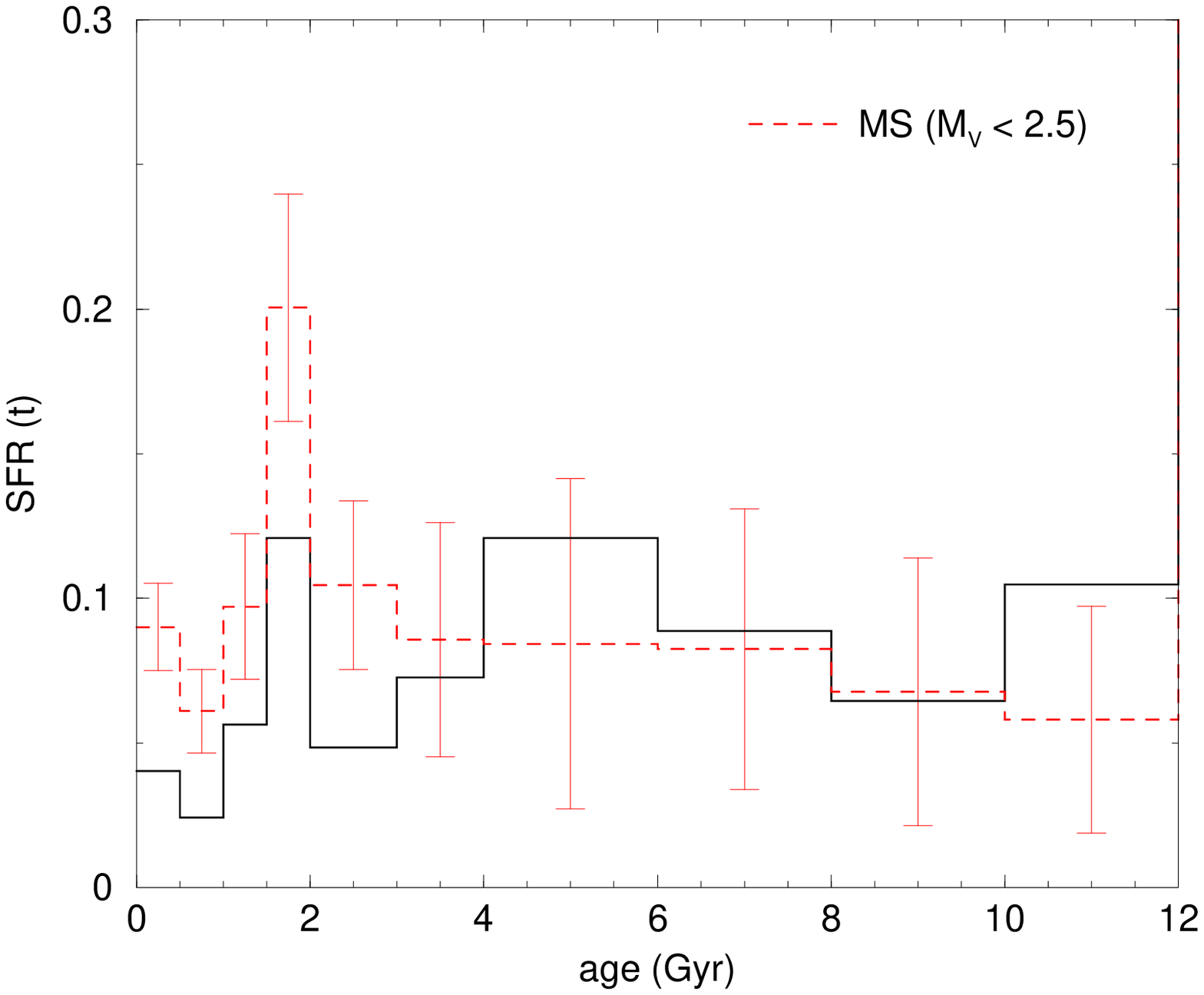}
\includegraphics[width=6.5cm,height=5.5cm]{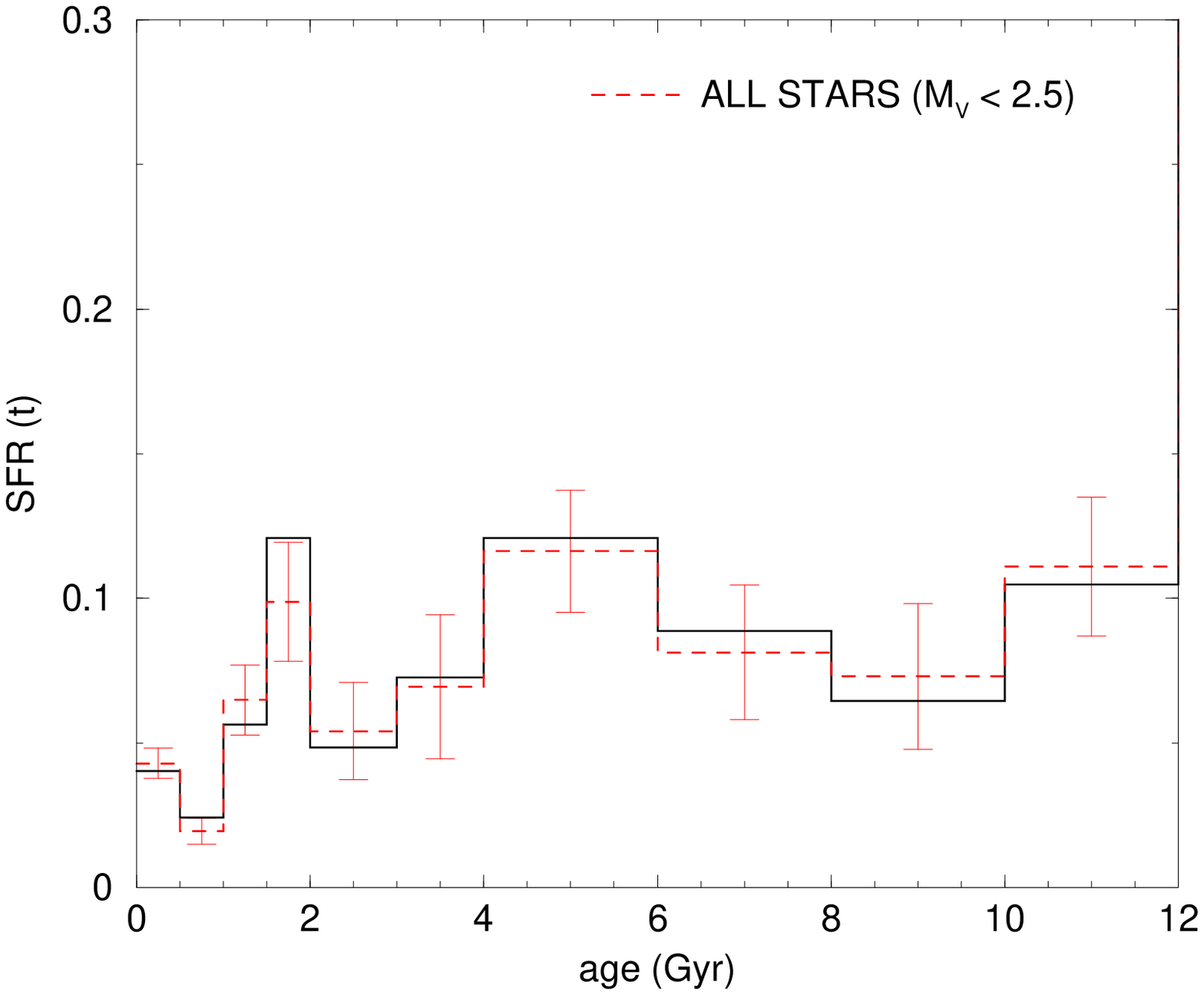}\\
\includegraphics[width=6.5cm,height=5.5cm]{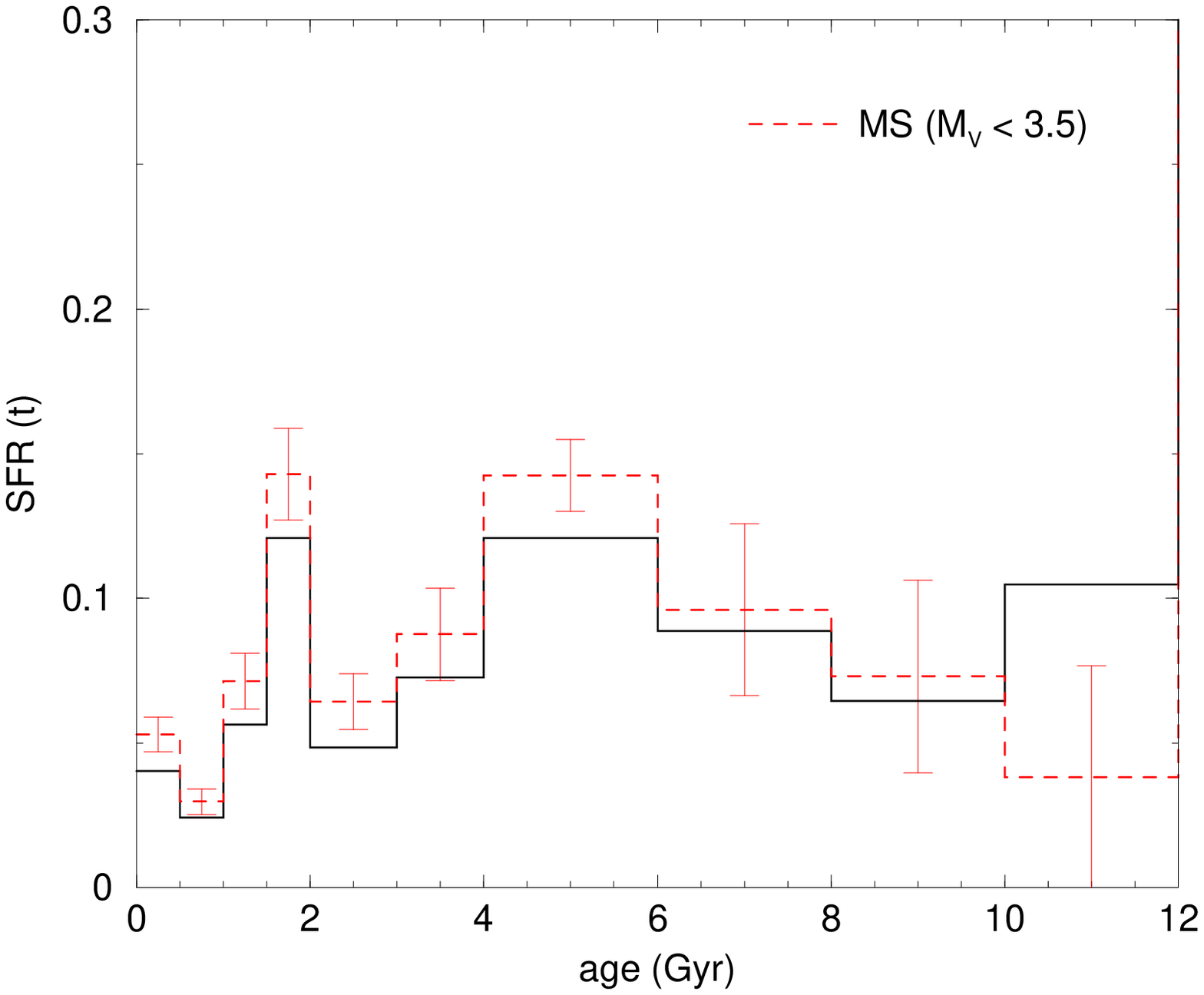}
\includegraphics[width=6.5cm,height=5.5cm]{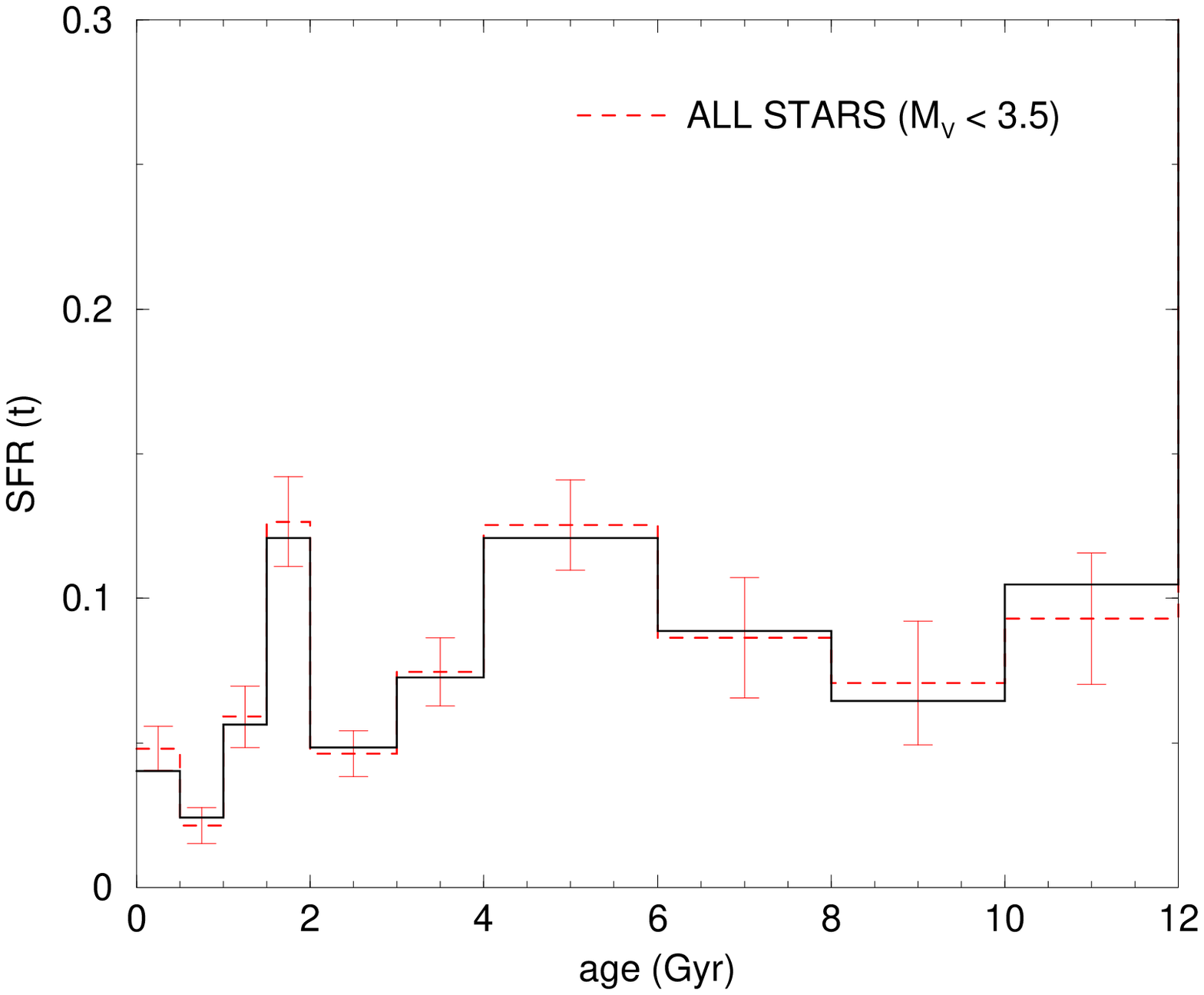}\\
\includegraphics[width=6.5cm,height=5.5cm]{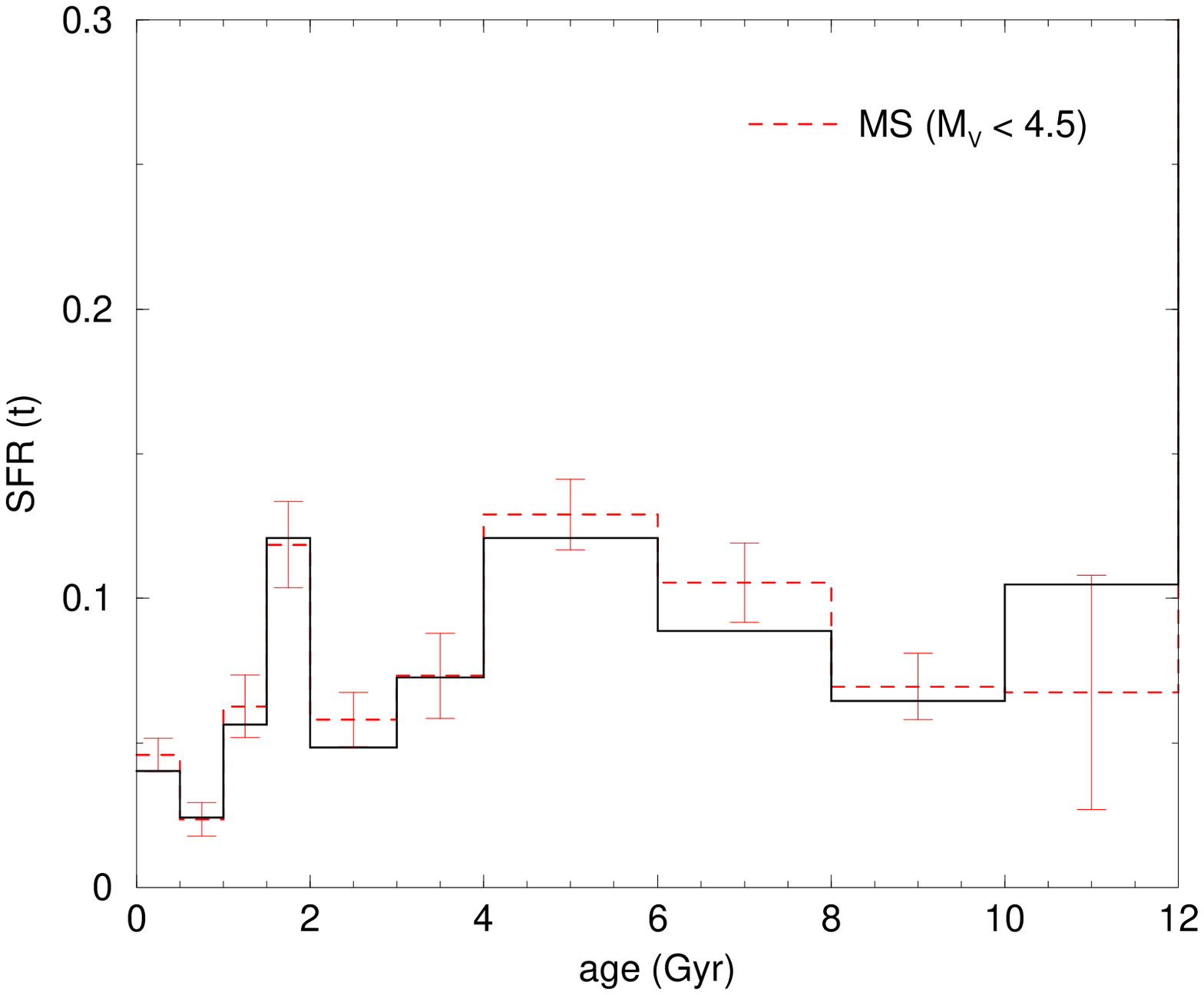}
\includegraphics[width=6.5cm,height=5.5cm]{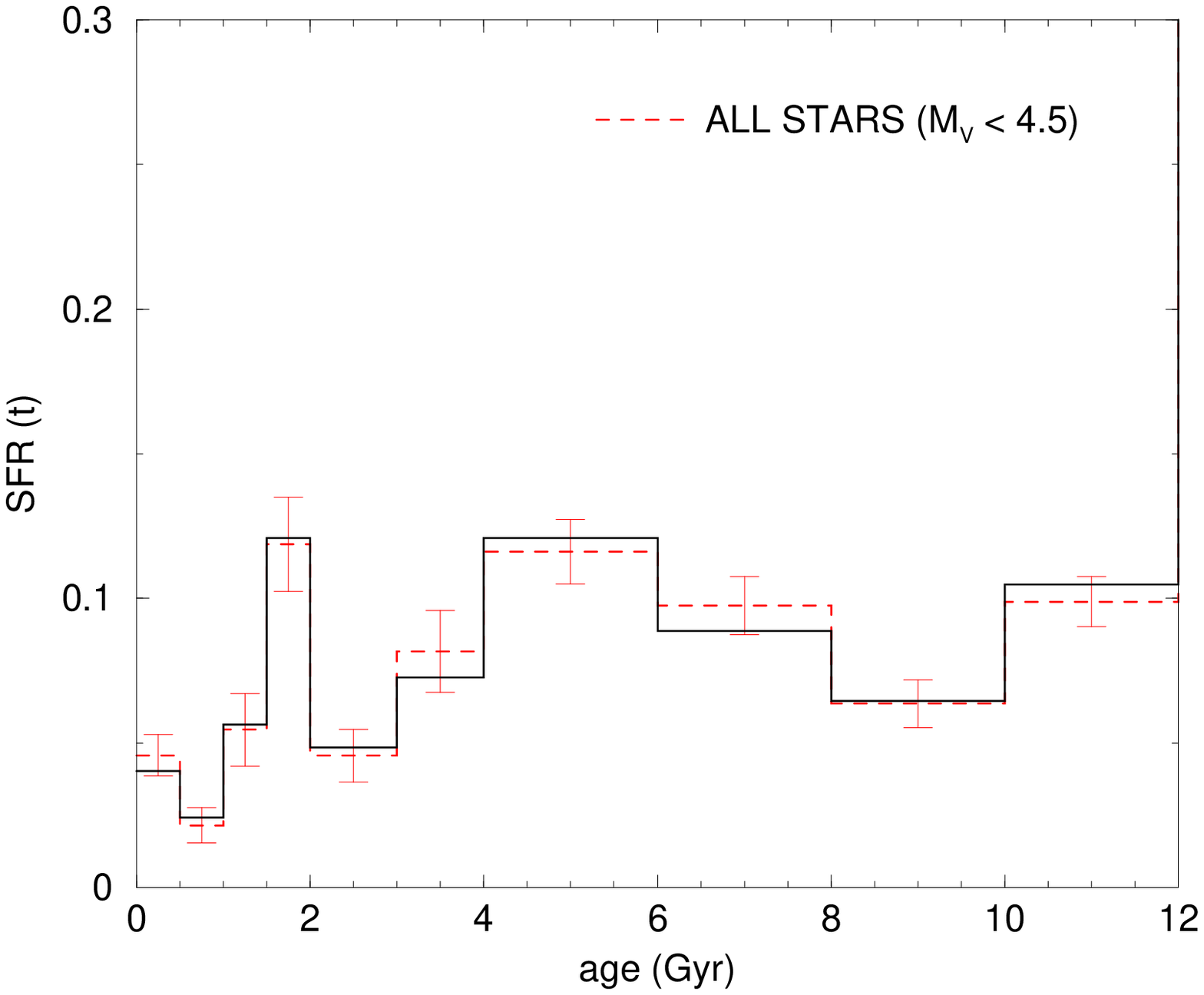}
\caption{``True'' (solid line) and recovered (dashed line) SFR from stars
  brighter than $M_V=2.5, 3.5, 4.5$. On the
  left the figures show the results obtained using only main sequence
  stars. The figures on
  the right represent the recovered SFR by including all the evolutionary
  phases.} 
\label{zone}
\end{figure*}

The completeness limit of the observed CMD limits our ability to exploiting
all the information contained in the CMD. In this section we explore how
different limits in absolute magnitude can modify the recovered SFR. We
analyze three different completeness limits: $M_{V}=2.5, 3.5, 4.5$. During the
test we separately select the contribution given by main sequence stars and
later evolutionary phases (RGB and red clump), defined as stars with
$B-V>0.8$.  The results are drawn in Fig. \ref{zone} (here and in the
following all SFR histograms have unit area).

Main sequence stars with $M_V < 2.5$ mag give information about
only the recent SFR, while we learn nothing about stars older than 3 Gyr (this
is evident from the large error bars for the recovered SFR beyond this age,
which mitigates the result and demonstrates the impossibility for the
procedure to recover the SFR due to the lack of information). The result is
significantly improved by also including later evolutionary phases. The
recovered SFR is close to the original one, even if the error bars are quite
large. This is due to the fact that one obtains information only from fast
evolutionary phases (clump and red giants for the past star formation and
upper main sequence for the recent one) and the probability of finding stars
in these zones is low. Obviously we are considering a perfect situation where
both the chemical composition and the IMF of the stars are well known so the
uncertainty is as minimum as possible; for real data other sources of
uncertainties occur.  Including stars to $M_{V}=3.5$ the precision of the
recovered SFR increases and from the main sequence alone one can obtain the
SFR up to $\sim 6$ Gyr. However, to study earlier epochs it is needed to
include later evolutionary phases. Also in this case the uncertainty in the
recovered SFR, for stars older than 6 Gyr, is large. The reason is the same:
for $M_V<3.5$, the information on the ancient star formation comes only from
late evolutionary phases which are too rapid to provide for a large number of
stars.  By increasing the magnitude limit at $M_{V}=4.5$, the entire SFR is
recovered with small uncertainties. The inclusion of the main sequence,
however, produces a systematic difference between the input and recovered SFR
for the oldest era while including late evolutionary phases leads to the right
solution.

\begin{figure}
\centering
\includegraphics[width=7.5cm]{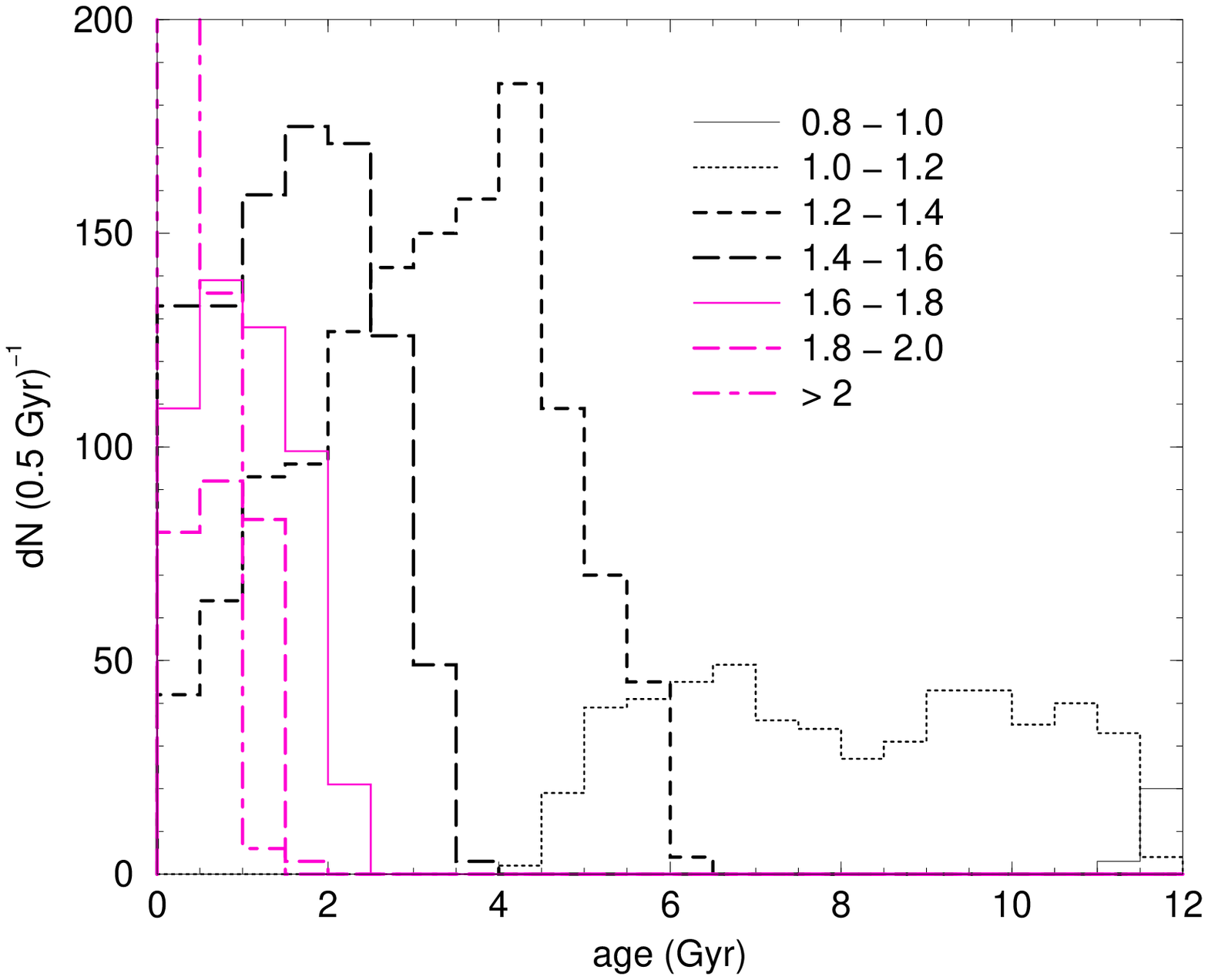}
\includegraphics[width=7.5cm]{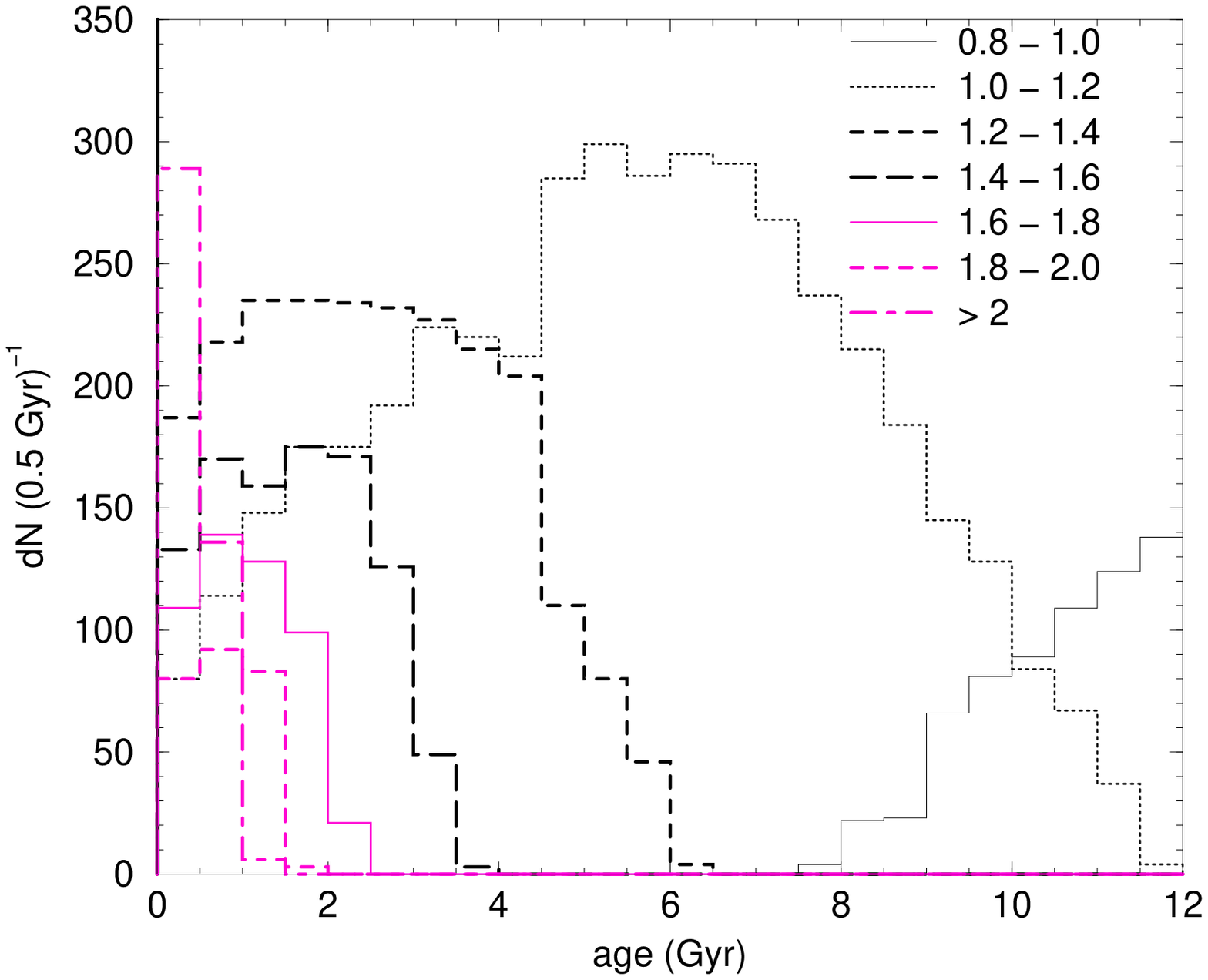}
\caption{Theoretical distributions in time of stars generated with a flat SFR (0-12 Gyr), Salpeter
IMF and solar composition. Different lines indicate different
mass ranges. In the upper panel only stars with visual absolute magnitude below 3.5
are plotted, while in the lower panel the magnitude limit is $M_{V}=4.5$.} 
\label{grotte} 
\end{figure}

\begin{figure}
\centering
\includegraphics[width=8cm]{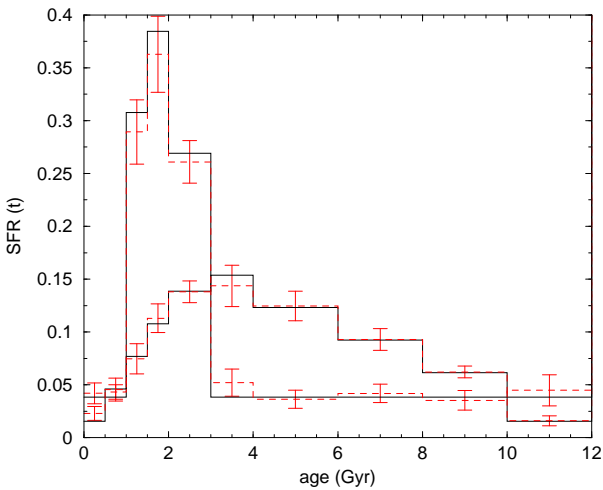}
\caption{As in Fig. \ref{zone} for $M_V<4.5$ and all evolutionary phases for
  different input SFR shapes.} 
\label{sfrs} 
\end{figure}

This finding is easier to understand 
if one looks how different masses contribute to different epoches of star
formation. Figures \ref{grotte} draw this map (for minimum luminosities
$M_{V}=3.5$ and $4.5$) for an artificial population built with a constant SFR
(0-12 Gyr) and Salpeter IMF.
If the minimum luminosity is set to $M_{V}= 3.5$ we cannot see the main sequence for masses
below $1.2\,M_{\odot}$. So this mass range the RGB and He burning stars provide information 
about the earlier epochs, 
but not on the recent SFR. At $M_{V}= 4.5$
we see the main sequence down to $1 \,M_{\odot}$, so we can analyze with this
mass range each period between now and $10-12$ Gyr ago. These results are \emph{not} linked to the particular SFR parameterization:
Fig. \ref{sfrs} shows the recovered SFR for different input SFRs.  

In the following we present test results involving artificial data adopting
the magnitude limit $M_V=4.5$.
\subsection{IMF - SFR degeneracy}
\label{IMF - SFR degeneracy}
Even if the form of the IMF is well defined for masses above $1\,M_{\odot}$
from observations and theoretical analyses (see e.g. Larson et al. \cite{lars}), the
precise value of the exponent is still debated. For instance, 
Kroupa (\cite{krou}) finds 
the Salpeter exponent 2.3$\pm$0.7 to be the most likely value, and any study 
of the local SFR must account for this uncertainty.

\begin{figure}
\centering
\includegraphics[width=8cm]{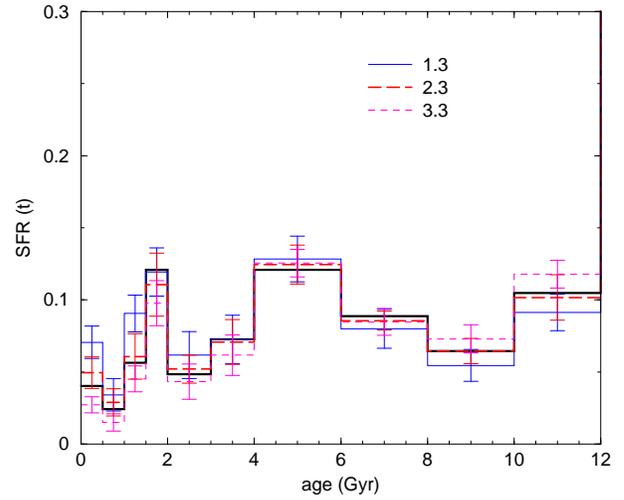}
\caption{The ``true'' (solid line) and the recovered SFR for
  different values (labeled) of the IMF exponent used in the artificial
  data. The IMF exponent used in the model is fixed to 2.3. } 
\label{imfs} 
\end{figure}

The possibility that changes in the IMF mimics the SFR effects on CMD is a
well known degeneracy. We analyzed the sensitivity of the recovered SFR to the
chosen IMF using different IMF exponents ($s=1.3,\, 2.3,\, 3.3,\, 4.3$) for
the artificial data while in the model a \emph{fixed IMF exponent equals to
2.3} was used.  The results are shown in Fig. \ref{imfs}.  It's noteworthy
that the input SFR is always recovered: even if a wrong IMF is adopted (that
is, different from the one used for the artificial data), it doesn't lead to a
wrong solution (at least for ``reasonable'' IMF exponents less than 4). In
conclusion, for the mass range covered by Hipparcos sample ($M_V\leq 4.5$),
\emph{the IMF exponent alone is not a crucial parameter if the AMR is known in
advance}. Figure \ref{grotte} shows this. For ages $ < 6$ Gyr the CMD is
populated by the whole mass spectrum (only very massive stars are dead), so a
variation of the IMF modifies the population in this age range in the same way
(the relative SFR is preserved). In contrast, the old eras (8 to 12 Gyr)
include only low mass objects (even intermediate mass stars are already
dead), thus the IMF variations mainly alters the ratio between old (older
than 8 Gyr) and recent star formation (see Fig. \ref{imfs}).

\subsection{Binaries - SFR degeneracy}
\label{Binaries - SFR degeneracy}
\begin{figure}
\centering \includegraphics[width=8cm,height=7cm]{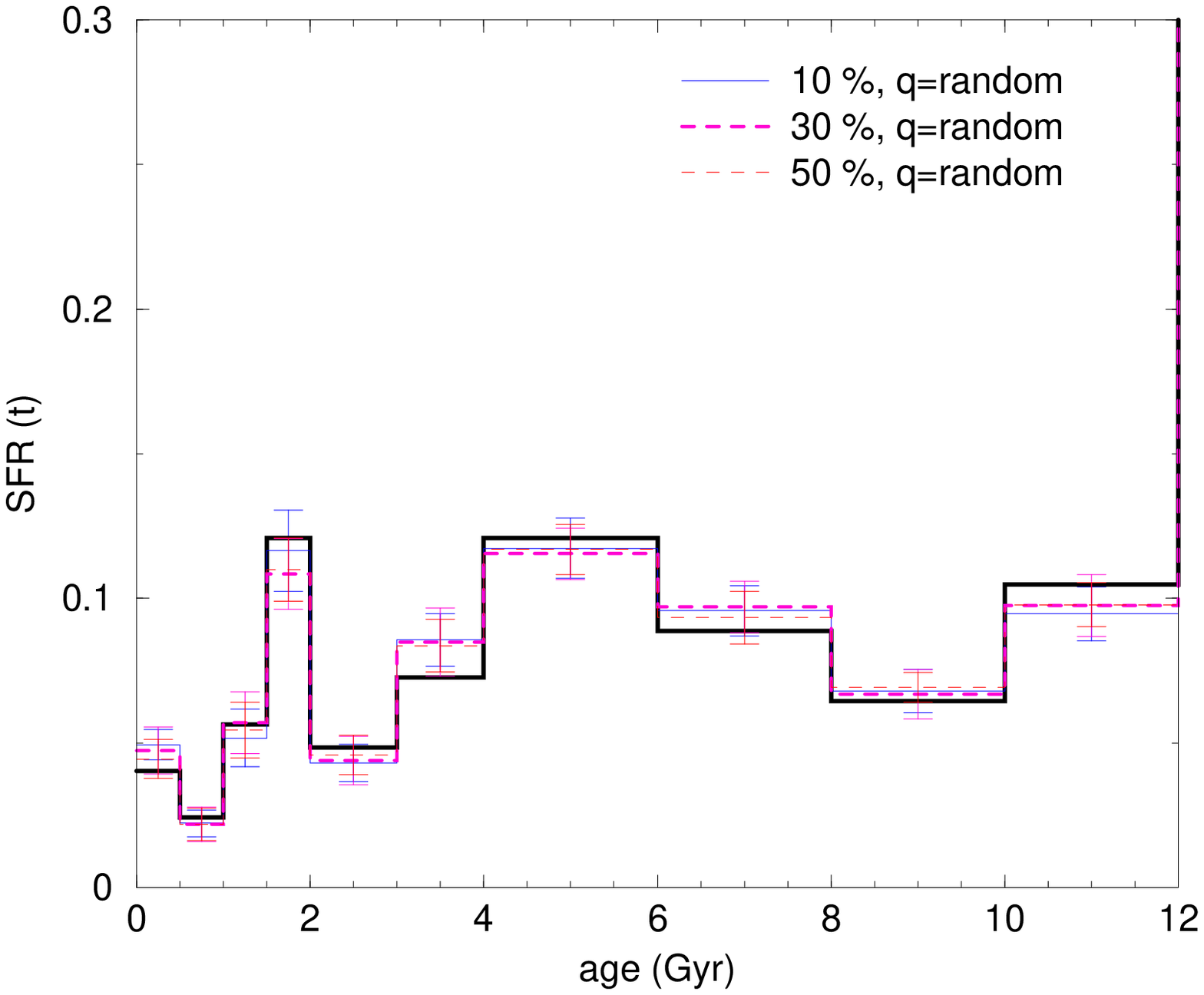}\\
\includegraphics[width=8cm,height=7cm]{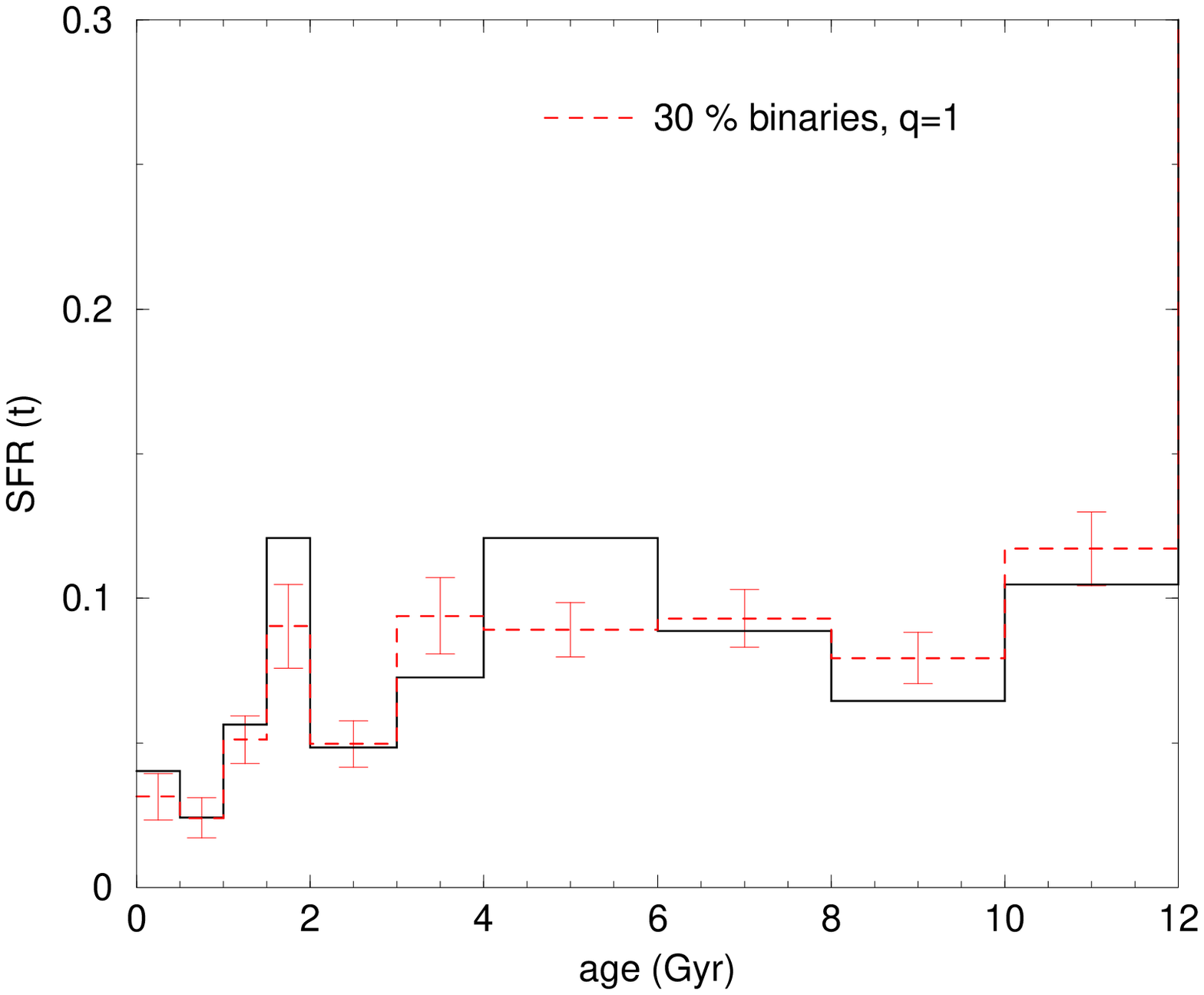}
\caption{``True'' (solid line) and recovered SFRs. The artificial
  data is generated with the indicated percentage of binaries and
  mass ratio (random number or unity), while the model is without binaries.}
\label{binarie} 
\end{figure}

Another source of uncertainty, when one looks at the solar neighborhood, is
the percentage of stars in unresolved binary systems. Our model doesn't
account for binary evolution with mass exchange: we assume that each binary
component evolves as a single star. Our knowledge of binary stars populations
and evolution in the local disk is far from perfect, so including interacting
binaries in the simulations would involve many unknown parameters (such as the
mass exchange rate, evolution of the separation, etc.). However, the mere
presence of a given percentage of unresolved binary systems affects the CMD
morphology. In order to perform this analysis, we have built the usual
artificial data using different prescriptions on the binary population (10\%,
30\%, 50\% of binaries with random and equal mass ratio). The partial CMDs
used in the model are built with the same composition and IMF adopted for the
artificial data, but without binaries.  The results of the simulations are
shown in figure \ref{binarie}. As found for the IMF, if the mass ratio is
uniformly distributed, we can recover the correct SFR independently on the
presence of binaries. In the extreme case of equal mass ratio ($q=1$), the
recovered SFR is severely biased.  In particular, the presence of binaries
doesn't affect the recent result, while the only modifications affect the old
SFR.

The explanation comes from the displacement that binaries cause from single
star CMD: if the mass ratio is uniform  the smearing effect is random and
the main sequence is merely wider (this effect is generally smaller than
0.05 mag, which is the binning size) and it looks like an increased photometric error 
for the recovered SFR.  If, however, the mass ratio is unity, the global
effect is systematic and the main sequence develops a parallel double: 
in this case the recovered SFR is systematically biased.  Thus, to 
recover a star formation rate, the choice of the
binary population is not crucial but if an extreme prescription is
adopted (e.g. equal masses), the recovered SFR may be biased.

\subsection{Metallicity - SFR degeneracy}

An old, metal-poor stellar populations can mimic
younger, metal-rich populations. Moreover, knowing that the solar
neighborhood is a mix of stars of different chemical compositions (see
e.g. Nordstr\"om et al. \cite{nord}), we expect a very complex influence of the
composition on the CMD morphology. These simple considerations oblige us to
check if it's still possible to recover the SFR when only a partial knowledge
on the age-metallicity relation is available.

 In these calculations we now assume that the IMF and the binary population are
the same both in the model and in the artificial data.

\subsubsection{Artificial data with single metallicity}\label{Artificial data with single metallicity}
\begin{figure}
\centering
\includegraphics[width=8cm]{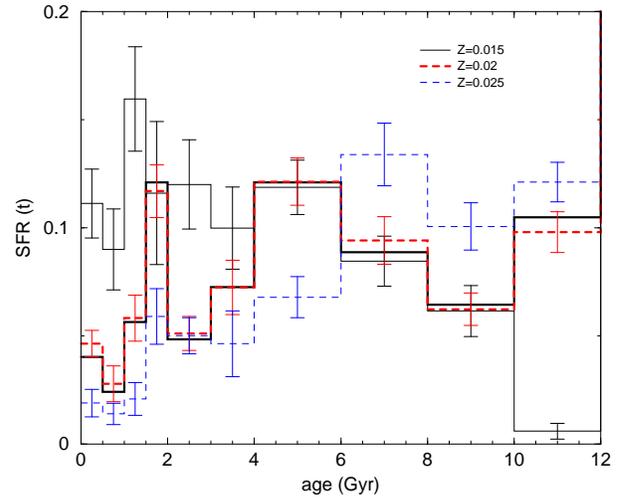}
\caption{Sensitivity test to metallicity. The adopted composition for the
model is solar. If the data have the same composition, the ``true'' SFR (heavy
solid line) is close to the recovered one (heavy dashed line). If the data metallicity is
slightly different from the solar value ($\Delta Z= \pm 0.005$), systematic
relevant deviations appear in the solution.}
\label{metawrong}
\end{figure}

\begin{figure*}
\centering
\includegraphics[width=8cm]{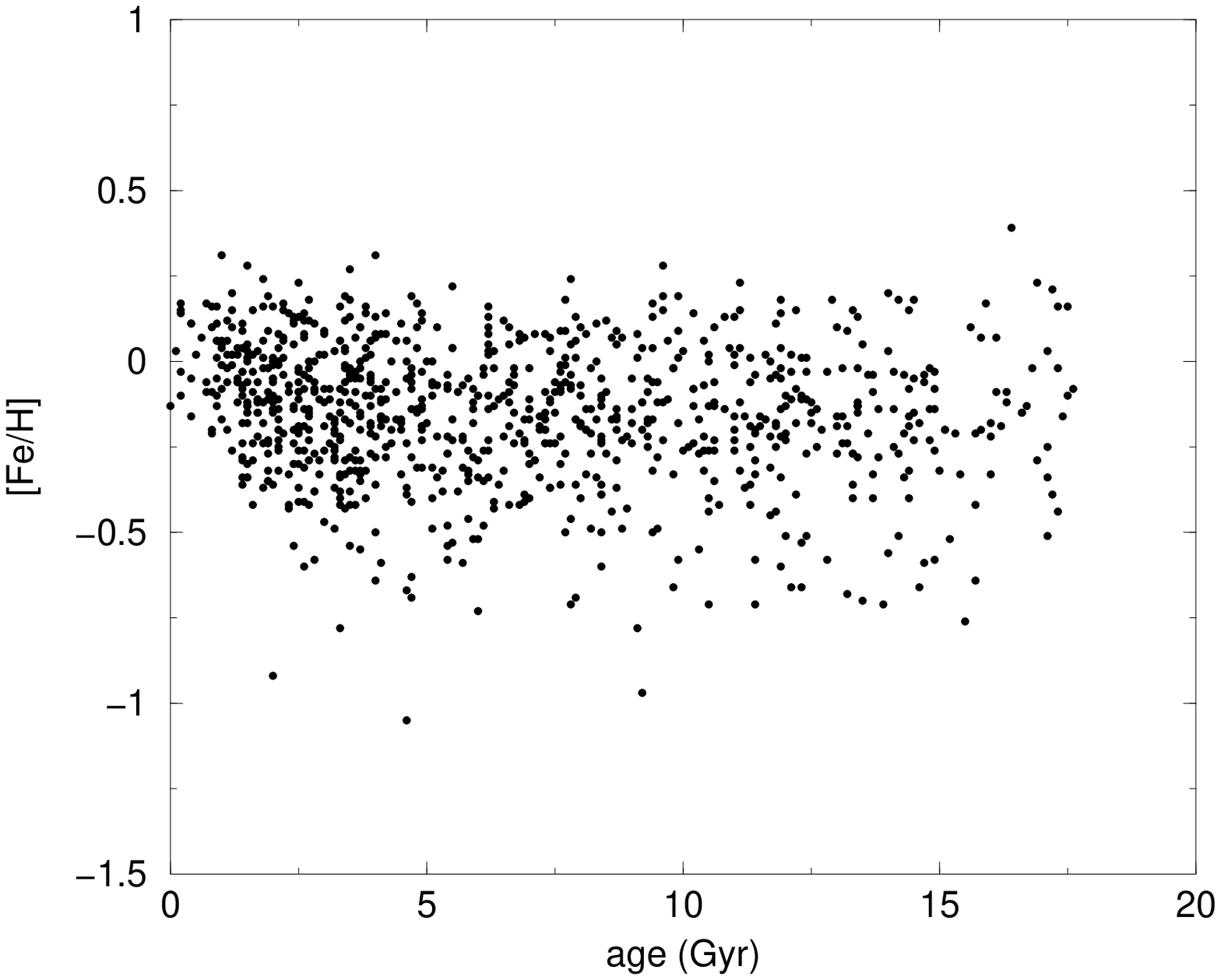}
\includegraphics[width=8cm]{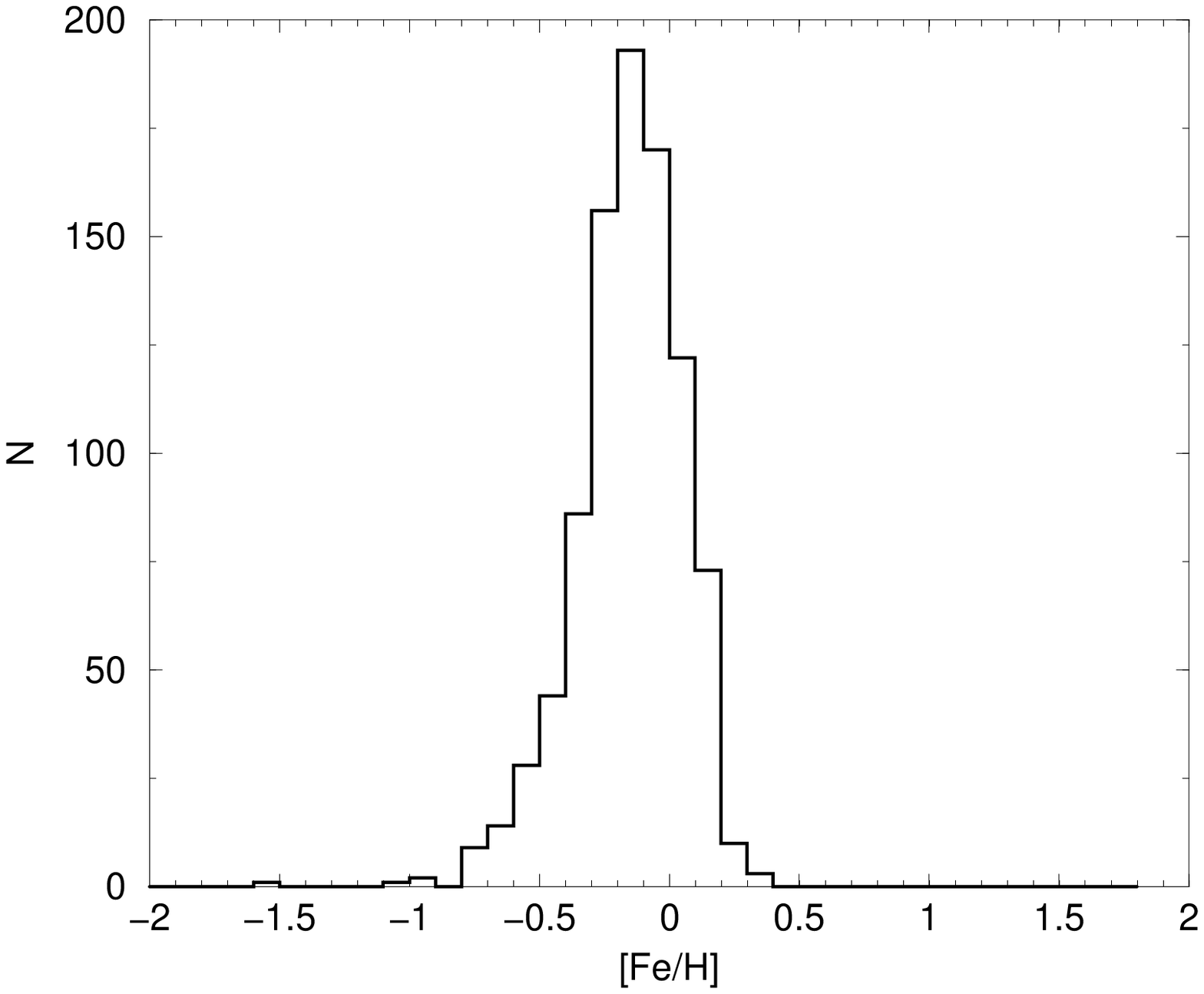}
\caption{Left panel: age-metallicity diagram for single stars (within 40 pc from Sun) with age determination
  better than 25\% (from Nordstr\"om et al. \cite{nord}). Right panel: the distribution of $[Fe/H]$ for
  the same sample.} 
\label{zt} 
\end{figure*}

As first step, we have built an artificial population with a single
metallicity (no AMR), then, the SFR is recovered assuming a wrong metallicity. 
The results are shown in Fig. \ref{metawrong}. 
If we adopt the same metallicity for the artificial data and for the model,
the SFR is recovered. \emph{However, if we slightly change the metallicity in
the data ($\Delta Z= \pm 0.005$), without changing the composition of the
model, systematic discrepancies appear in the recovered SFR}.  If the
artificial data are metal poor compared to the model, the recovered old SFR is
underestimated (and the recent one is overestimated) and oppositely for metal
richer artificial data.  This result is a strong warning about the widely used
assumption of solar composition for nearby stars: \emph{small deviations from
the solar value could bias the derived SFR}. Moreover, we know that deviations
from solar values exist and are usually much larger than $\Delta Z= \pm
0.005$.  Figure \ref{zt} shows the age-metallicity relation and the
metallicity distribution by Nordstr\"om et al. (\cite{nord}), the most
representative census for ages and metallicities in solar neighborhood. This
is characterized by a constant mean metallicity and a large scatter at all
ages (about $\sigma \sim 0.2\,\rm dex$ in $[Fe/H]$). The quoted formal error
of $\sim 0.1$ dex in $[Fe/H]$ cannot account for the observed spread.

\subsubsection{Artificial data with an age-metallicity dispersion}
In order to test the
sensitivity to a metallicity dispersion, artificial data were generated with
the observed mean metallicity plus a variable dispersion (the explored range
is from $\sigma=0.01$ dex to $\sigma=0.2$ dex in $[Fe/H]$).  The conversion
between $[Fe/H]$ and $Z$ that is appropriate for our models calculated for the
GN93 composition ($[Z/X]_{\odot}=0.0245$) is:
\begin{eqnarray}
 \log Z = 0.73\times 10^{([Fe/H]-1.61)}
\label{fehz} 
\end{eqnarray}
The enhancement of $\alpha$ elements is not included because it has shown to be negligible for disk stars.

\begin{figure*}
\centering
\includegraphics[width=7cm]{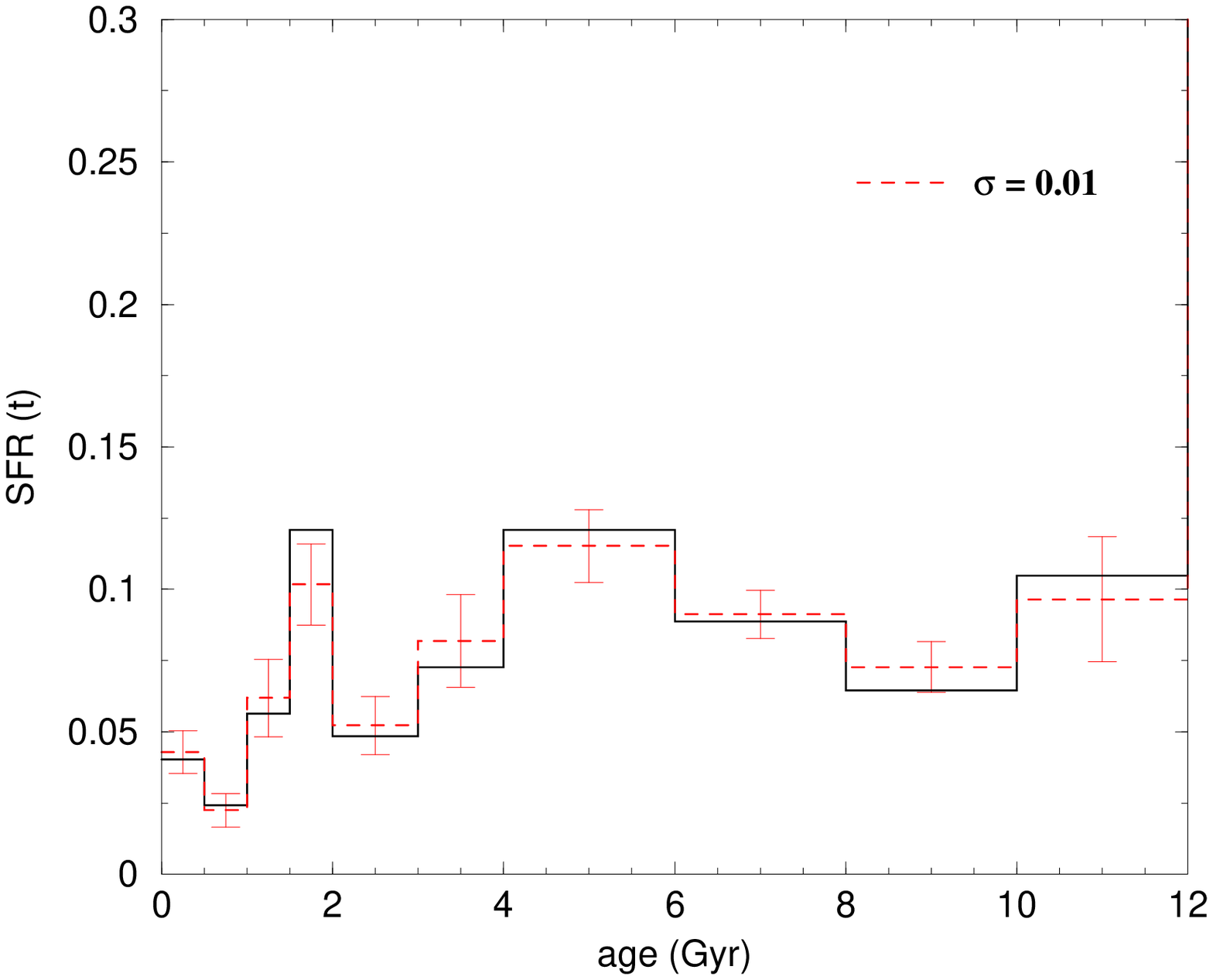}
\includegraphics[width=7cm]{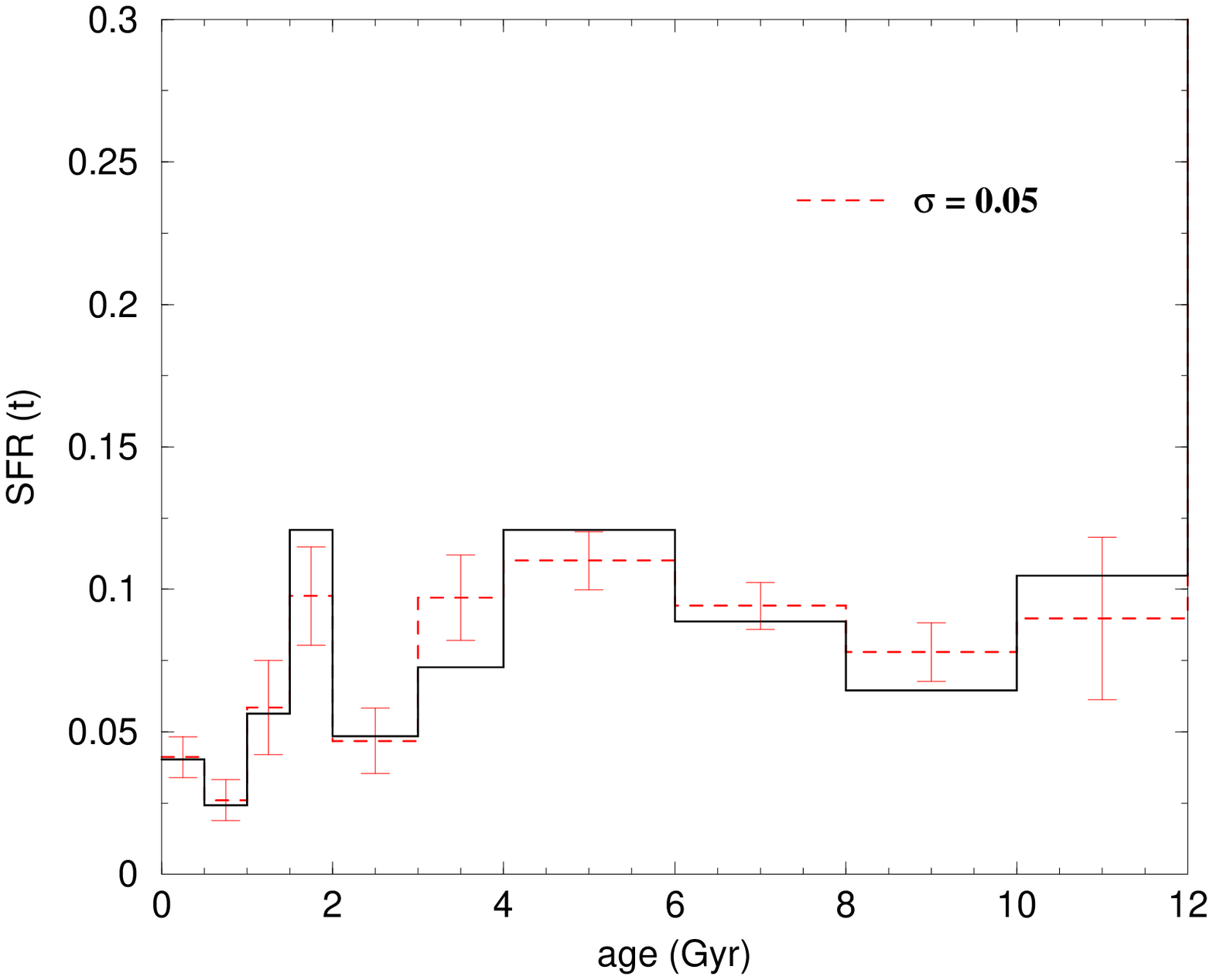}\\
\includegraphics[width=7cm]{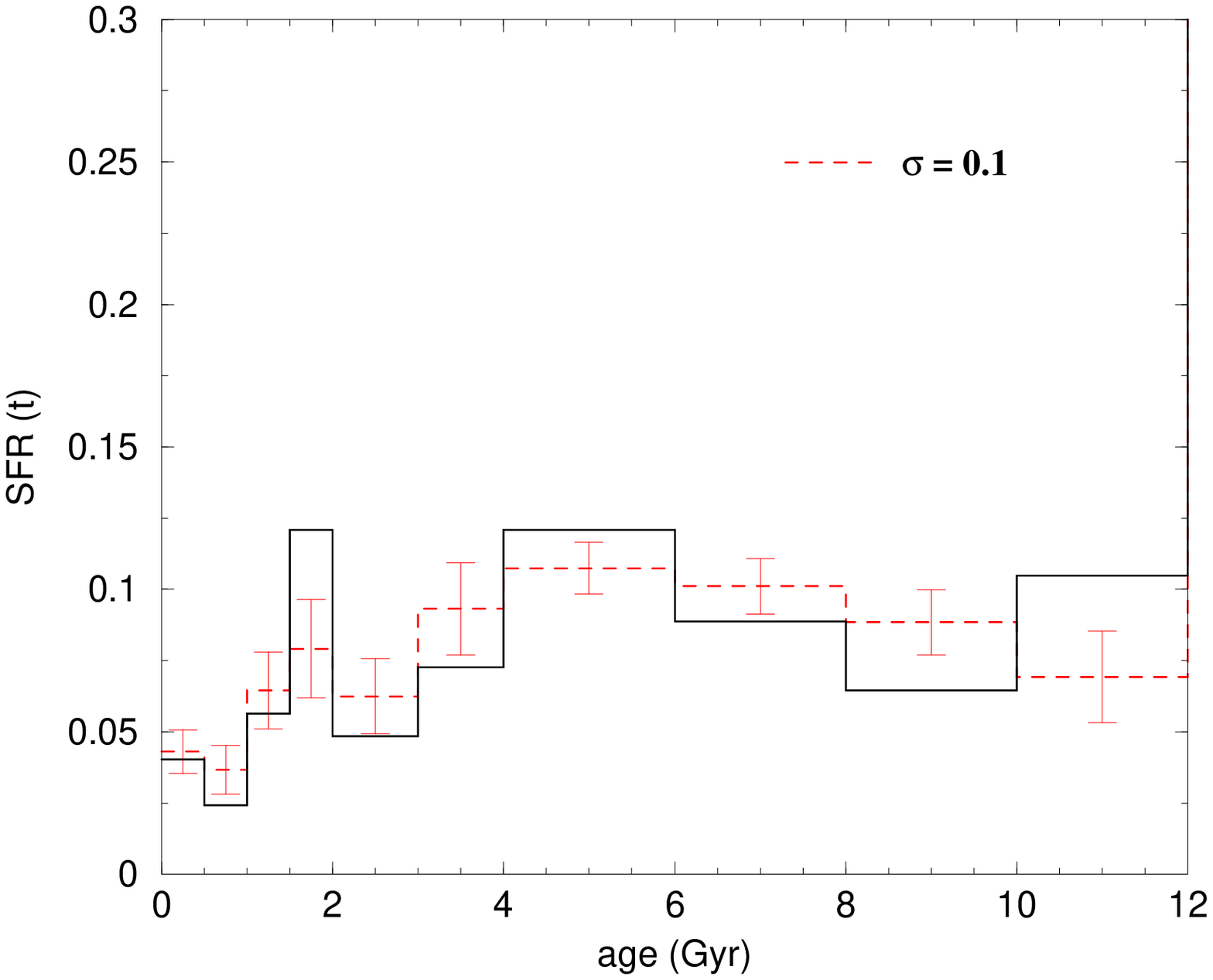}
\includegraphics[width=7cm]{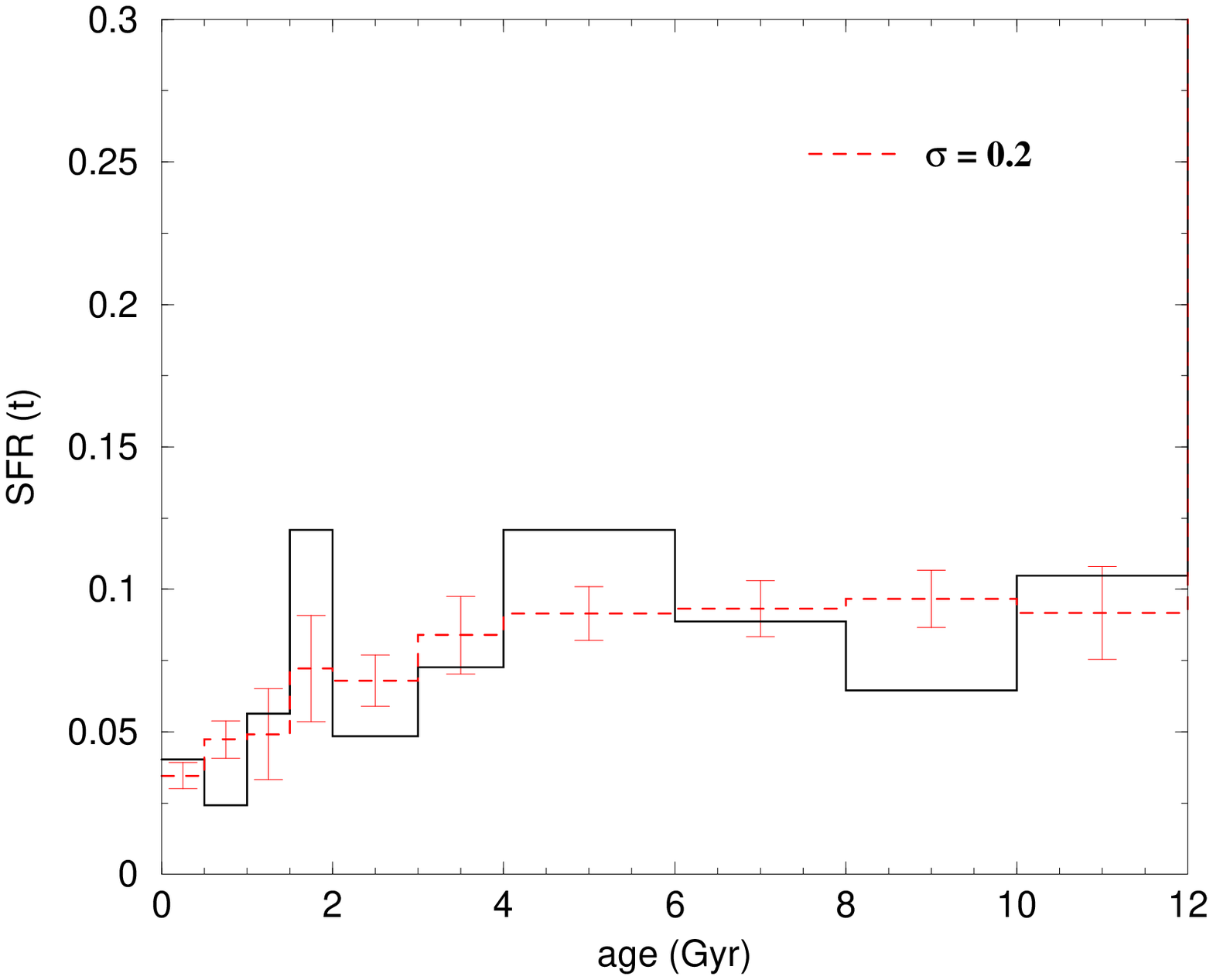}
\caption{Solid line: SFR assumed for the artificial data. Dashed line: recovered SFR. The $\sigma$ value
  indicates the dispersion in $[Fe/H]$, used for the artificial data. The
  model has the same mean metallicity, but no dispersion.} 
\label{tantedisp} 
\end{figure*}

The SFR was searched,
adopting in the model the same mean metallicity of the artificial data, but
\emph{without metallicity spread}. The results are shown in Fig. \ref{tantedisp}, 
where the solid line is the input SFR and the dashed line the recovered one.
Above $\sigma=0.1$ dex, most of the information contained in the SFR is lost: this numerical experiment has pointed out how the dispersion in
metallicity can be a critical factor. \emph{A wrong estimate of the dispersion
leads to a wrong solution.}

For the final test we adopt the same dispersion both in the artificial data
and in the model. The idea is to check if in presence of dispersion, even when we
rightly estimate the metallicity dispersion of the data, the spread of the
CMD due to the metallicity spread still allows to recover the underlying SFR.
In order to do a realistic attempt, we adopt the metallicity distribution by
Nordstr\"om et al. (\cite{nord}) (Fig. \ref{zt}).

\begin{figure}
\centering
\includegraphics[width=8cm]{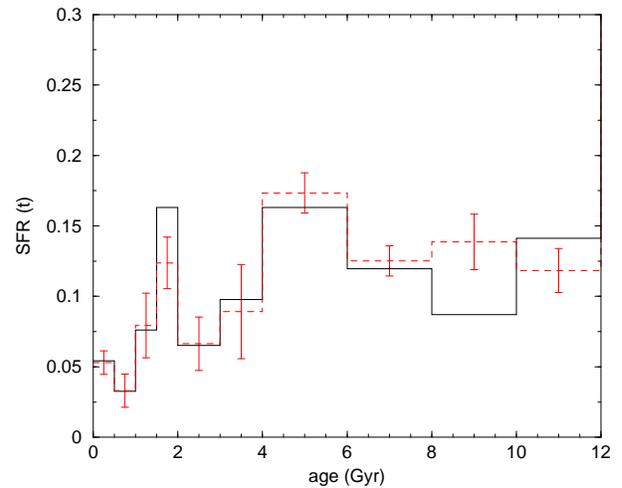}
\caption{Input SFR (solid line) compared with the recovered SFR (dashed line). 
Model and artificial data have the same metallicity dispersion. }
\label{halfrec}
\end{figure}

The overall effect is a broadening of the
partials CMDs. 
The SFR extraction is presented in figure \ref{halfrec} for an artificial population generated with this AMR.
There are systematic shifts between the recovered and the input SFRs,
indicating a limit in our ability to distinguish different stages of star
formation (for a comparison with the single metallicity tests, see section
~\ref{Artificial data with single metallicity}), but the trend is
preserved. The implication for real data is actually encouraging: if the
nearby stars show an age-metallicity relation like the N\"ordstrom et
al. result, the application of the model to Hipparcos stars can give
information on the real SFR.

\subsection{Contamination from clusters and associations}
The solar neighborhood includes 
stellar clusters or part of them. We have removed about 80 stars (mainly
Hyades stars) within 80 pc and brighter than $M_V=3.5$. This number may appear
small ($\sim 2\%$ of the total sample), but these object are concentrated in
time so they can produce a burst in the SFR (at $\sim 0.5$ Gyr, for the
Hyades) which does not represent a galactic field property.

\begin{figure}
\centering
\includegraphics[width=8cm]{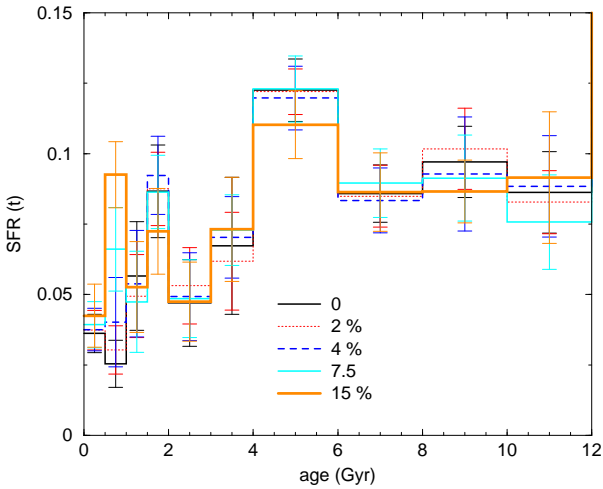}
\caption{Recovered star formation rates for different contaminations (the
  percentage is labeled) of Hyades-like stars. }
\label{clusterhip}
\end{figure}

\begin{figure}
\centering
\includegraphics[width=8cm]{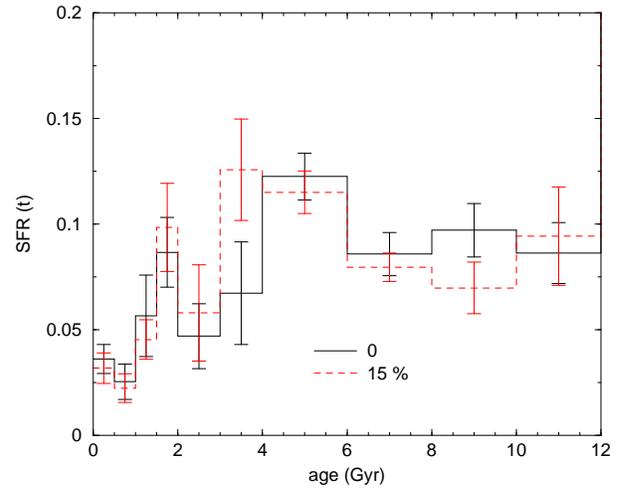}
\caption{Solid line and dashed line are respectively the recovered SFR when the
  contamination by cluster stars (2 Gyr old) is 0 and 15\%. }
\label{clusterhip2}
\end{figure}

 However, we cannot exclude that some cluster members remain as yet
 unidentified, so an interesting analysis involves the cluster impact on the
 recovered SFR. 
We have contaminated an artificial sample with a variable percentage
of synthetic Hyades-like stars (500 Myr and solar metallicity), from 2\% to
15\%, with the results shown in Fig. \ref{clusterhip}. At 2\% contamination the
SFR changes within the error bars. Increasing the cluster percentage, the
peak at 500 Myr becomes progressively more evident. At 15\%, the recovered SFR is
perturbed on a scale of 5 Gyr. 
The same test has been performed with a synthetic cluster at 2 Gyr. Figure
\ref{clusterhip2} shows the recovered SFR when 15\% of synthetic cluster stars
is added to the artificial data. It is evident that the changes in the
recovered SFR are not isolated at 2 Gyr, but the whole SFR shape between 2 Gyr
and 7 Gyr is altered.

\section{Comparison with real data}
 
We are now ready to present the SFR extraction from the real Hipparcos
data. Because both the IMF and the binary population are not crucial factors
(see sections ~\ref{IMF - SFR degeneracy} and ~\ref{Binaries - SFR
degeneracy}), we have fixed them: the IMF exponent is Salpeter, there are no
binaries, and we have again assumed the Nordstr\"om et al. AMR. Before applying
the SFR extraction method to real data, we must treat the observational
errors. Cignoni \& Shore (2006) show how the Richardson-Lucy algorithm allows to
restore the original CMD corrupted by a point spread function. Here, we apply the SFR extraction to the Hipparcos CMD that was previously deconvolved
by this algorithm. However, the result of a R-L restoration is a two
dimensional histogram and the information on the single stars is lost. Thus we
cannot directly apply the bootstrap technique to determine the variance on the
recovered SFR. A trick to avoid this problem is to construct bootstrap
replicates of the data \emph{before the Richardson Lucy restoration}. Then
each bootstrap data is reconstructed with the R-L algorithm and the SFR is
obtained for each reconstruction.  This set furnishes the mean and variance
for the final SFR.
 
\begin{figure}
\centering
\includegraphics[width=7cm]{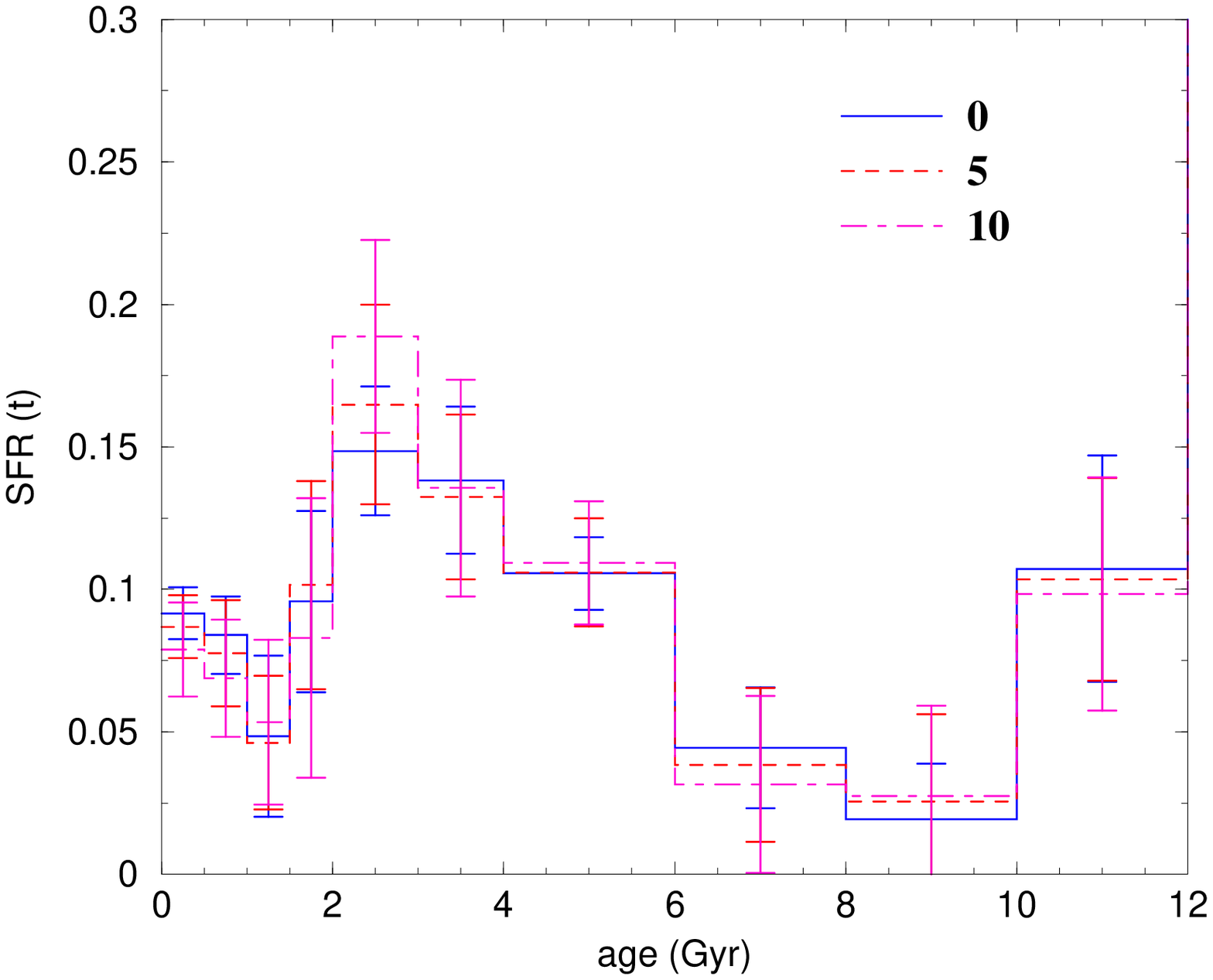}\\
\includegraphics[width=7cm]{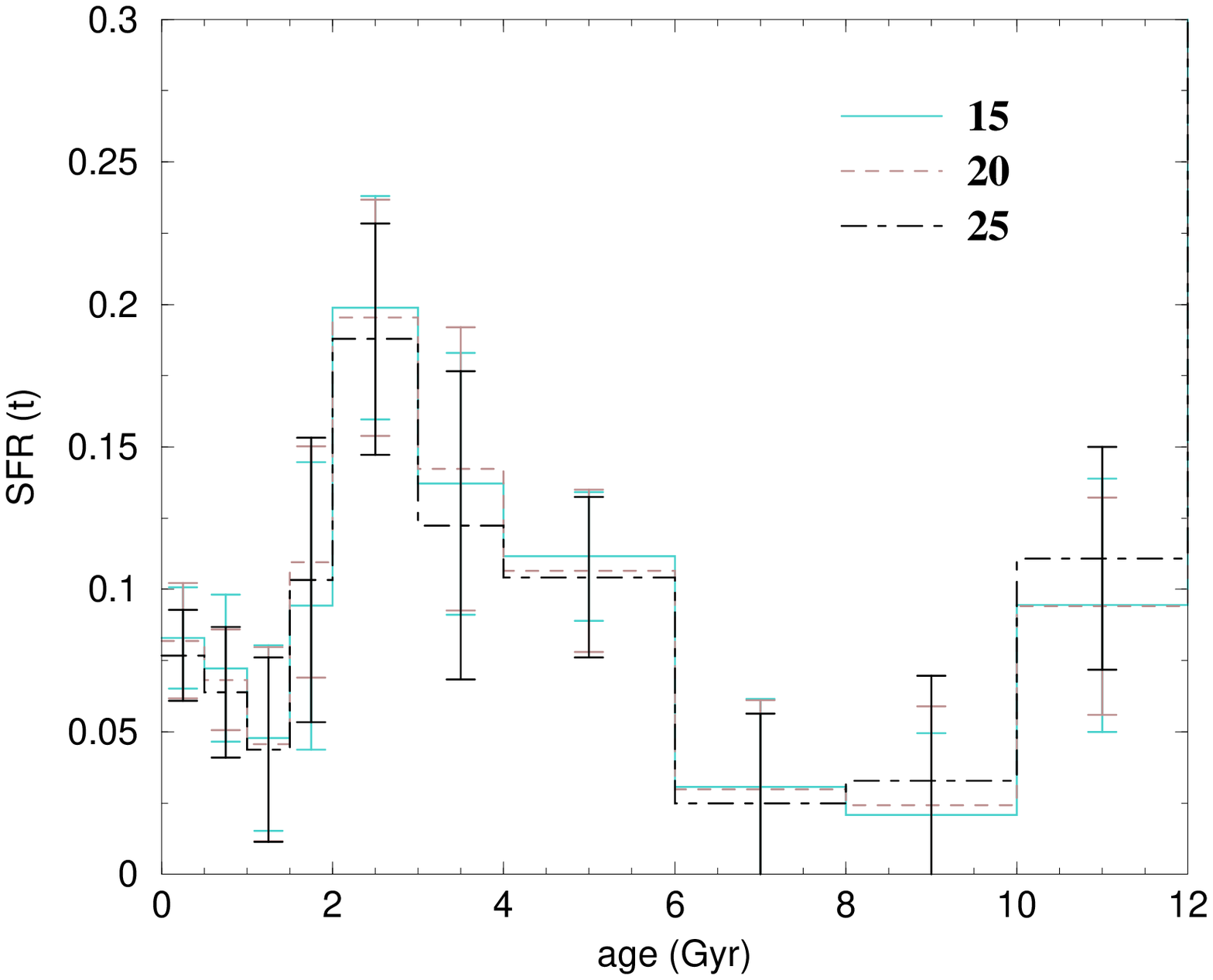}
\caption{The SFR recovered from the Hipparcos sample. The comparison area
  involves all stars brighter than $M_V=3.5$. Different
  lines show the result after the labeled number of R-L iterations.}
\label{lucy_sfr}
\end{figure}

The R-L algorithm is performed with the PSF built from the observational
absolute magnitude error distribution. Figure \ref{lucy_sfr} shows the
results: the different curves represent the recovered SFR, after respectively
5, 10, 15, 20, 25 R-L restorations. In order to avoid artifacts, the
restoration is stopped at the 25-th iteration when the bulk of the restoration
is done (see discussion in Cignoni \& Shore \cite{cign}). For comparison, Fig. \ref{lucy_sfr} (upper panel) shows the recovered SFR
(labeled with 0) when our method is applied to the data without R-L
restorations.

The global effect of the restorations is small \footnote{This means that the
uncertainties in the Hipparcos data (at the luminosities of our sample) are
small. Cignoni \& Shore (2006) find that this reconstruction is especially useful
when the uncertainties are much more larger.} and it is most evident around
2-3 Gyr. Over the 10th iteration, the solution is very stable and the only
change is the increase of the estimated uncertainties because of the noise
amplification.

Before discussing the recovered SFR, it is necessary to recall that the
implemented Nordstr\"om et al. AMR is uncertain for ages lower than 1.5 Gyr and
greater than 7 Gyr, thus the result for these ages could be unavoidably biased. Moreover, all the
information (at $M_V<3.5$) on the SFR older than 7
Gyr comes from evolved stars, with the associated under-population problems. All
these points will be discussed further. For the moment, we describe the
results for the SFR:
\begin{itemize}
\item {A bump in the time interval 10-12 Gyr;}
\item {A modest activity in the time interval 7-10 Gyr;}
\item {A steep increase from 2 Gyr to 6 Gyr;}
\item {A modest activity during the last 1 Gyr.}
\end{itemize}

\begin{figure}
\centering
\includegraphics[width=7cm]{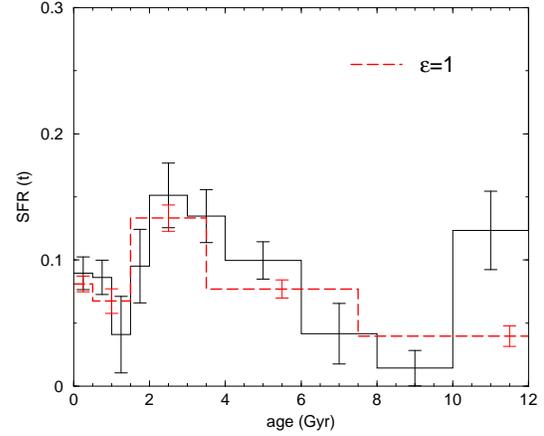}\\
\caption{The recovered SFR for a power law step resolution (dashed line;
  $\varepsilon=1$, see text) is compared with the result of
  Fig. \ref{lucy_sfr}. Both the SFR are obtained after 1 R-L restoration.}
\label{nuova1}
\end{figure}

In order to check the robustness of our finding, we tried to recover the SFR
by adopting a different time resolution. In particular we tested a power law
stepping $\delta t = [{(1+\varepsilon)^n}\,\times 0.5\,Gyr]$, with $\delta t$
the step duration, $\varepsilon$ a parameter and $n$ a running index (integer, starting from 0). Figure \ref{nuova1} shows the recovered SFR for $\varepsilon=1$.
The coarser time steps allow to reduce the uncertainties, but the overall
shape for ages less than 8 Gyr is confirmed. As we will discuss later, the results for greater ages are affected by many sources of
uncertainty. In the following, the temporal resolution will be fixed to the
prescription of section \ref{stepping}. Moreover the number of R-L restoration
is fixed at 15.

\subsection{Warnings about the AMR}
\label{nordmeth}
Our SFR represents the most probable result provided that the model is not
 biased. As already discussed, the observed AMR is chosen mainly
 because it arises from a very wide observational sample. However, it is still
 affected by three important biases: 
\begin{enumerate}
\item it was built looking for F-G type stars. This selection
 was done choosing stars between suitable blue and red color boundaries
 (by means of $(b-y)$ Str\"omgren color, which is almost metallicity
 independent). However, as a consequence of the blue cut-off, the younger
 metal poor stars are under-represented in the final AMR;

 \item Due to the observational errors, the stellar age determination is
 progressively more and more difficult as a star is close to the zero age main
 sequence (where the stellar tracks degenerate). Consequently, the
 Nordstr\"om AMR is poorly known for very young stars and the AMR we used below
 1 Gyr is an extrapolation;

\item The age determination is also a problem for
 stars older than 8 Gyr, because at these ages the main sequence is populated
 by low
 mass stars which evolve in a restricted region of the CMD. As a consequence,
 the ancient part of the AMR is
 given with very large uncertainties in the age (e.g. the presence of stars 
 older than 13 Gyr). 
\end{enumerate}

\begin{figure}
\centering
\includegraphics[width=7cm]{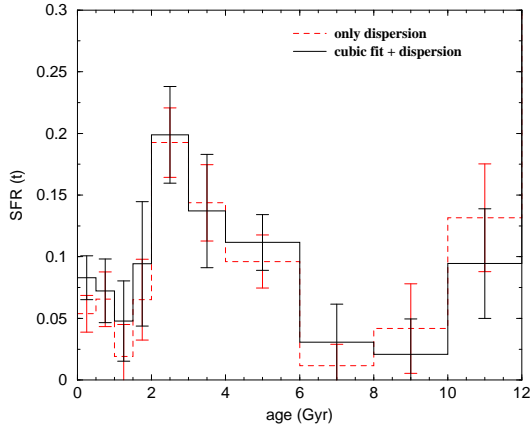}
\caption{Dependence of the result on a different parameterization of the
  observed AMR. The dashed line indicates the SFR that is obtained
  implementing the metallicity spread of the Nordstr\"om et al. AMR. The heavy
  line indicates the resulting SFR when a cubic interpolation of the
  Nordstr\"om et al. data plus the same observed dispersion is adopted, see
  text.}
\label{cubicflat}
\end{figure}

Although we have selected stars with a relative uncertainty for the age better
than 25\%, the previous points lead to doubts about the recovered SFR for
stars younger than 1 Gyr and older than 8 Gyr. In particular, the recovered
activity during the last 1 Gyr is partially due to the way the AMR is
parameterized in our model. Figure \ref{cubicflat} shows the effect of a
different parameterization: the solid line is the resulting SFR if the adopted AMR is
a polynomial fit (cubic) plus the dispersion, the dashed line is the result when the dispersion alone is implemented.

\subsection{Warning about the adopted completeness limit}
\begin{figure}
\centering
\includegraphics[width=7cm]{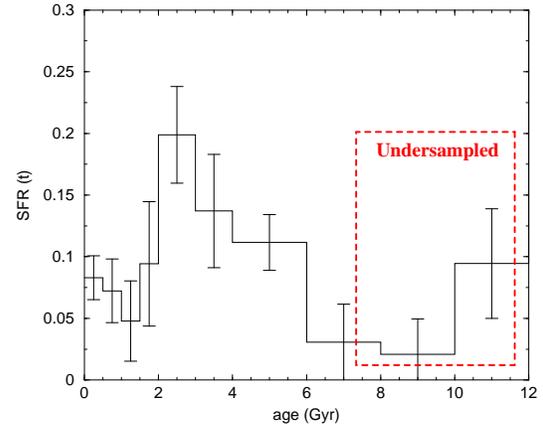}
\caption{The highlighted region identifies the time interval where the recovered SFR
  is undersampled (because of the magnitude cut at $M_V=3.5$, the only tracers
  at these ages are red giant and helium clump stars).}
\label{unders}
\end{figure}

Another point is the completeness limit $M_V=3.5$: in section 5 we
showed how the information on the old SFR is enough limited by the
completeness limit, with the full information only available with an
hypothetical sample complete up to $M_V=4.5$. With the $M_V=3.5$ cut-off, the
only tracers of the star formation older than 7 Gyr (see Fig. \ref{grotte})
are red giants, clump stars and subgiants. For this reason, at ages older than
7 Gyr the recovered SFR is certainly undersampled (see Fig. \ref{unders}). Thus, the bump between 10 and 12 Gyr could be an artifact.
Only a deeper volume limited sample will provide a better
understanding on the old SFR.

\section{Kinematical selection}
 A genuine star formation rate
 should represent the number of stars born at each time in our
 volume; this condition can fall, for example when: 
 \begin{itemize}
\item[1)] Old disk stars may have \emph{diffused} into a larger volume, so the
  old local SFR may be undersampled: the stellar velocities are randomized
  through chance encounters with interstellar clouds, they gain energy and
  increase the velocity dispersion.

\item[2)] ``Hot'' populations may contaminate the sample. Thick disk and
  halo stars have kinematical properties that could have been fixed before the
  disk developed. These stars sample a much larger volume of the disk
  stars and they are weakly represented in the solar neighborhood. 
  \end{itemize}

In these cases, the recovered SFR would be a mere census of the ages of the
stars actually present in the
solar neighborhood. In order to avoid thick disk/halo contaminations and
to check the amount of orbital diffusion of old disk stars, we have evaluated the
Galactic velocity components for all stars in the sample and recovered the SFR
for different kinematically selected subsamples.

\begin{figure}
\centering
\includegraphics[width=7cm]{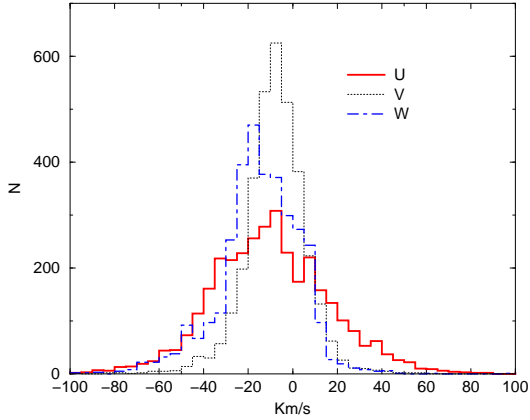}
\caption{Distribution of U,V and W velocities (referred to the LSR) for all
  stars in the sample with a measured radial velocity and $M_{V}<3.5$. }
\label{velox}
\end{figure}

\begin{figure}
\centering
\includegraphics[width=7cm]{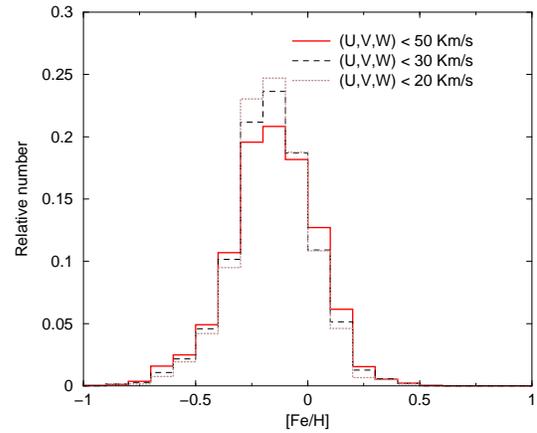}
\caption{Normalized $[Fe/H]$ distributions (data by Nordstr\"om et al. \cite{nord}) for stars
  with the labeled kinematic selection. }
\label{fehk}
\end{figure}

The Hipparcos mission measured proper motions which, together with the
parallaxes, give tangential velocities $V_T$. For most of the stars in our
sample, a measured radial velocity is available (from the SIMBAD
database \footnote{SIMBAD database is available at the following URL:
http://simbad.u-strasbg.fr/Simbad}). With these data, we have computed the
Galactic velocities U,V and W for more than 90\% of the stars in the sample,
corrected for the solar motion relative to the Local Standard of Rest (LSR,
$U_{\odot}=+10.0$ Km/s, $V_{\odot}=+5.2$ Km/s, $W_{\odot}=+7.2$ Km/s according
to Dehnen \& Binney \cite{deh}). Figure \ref{velox} shows the distribution of U,V
and W velocities for all stars in the sample with measured radial velocities.
In order to search for stars with thin disk properties we need a kinematic
criteria. Table \ref{tab} summarizes recent results for the kinematic
properties of the thin disk, thick disk and halo (values from Bensby et
al. \cite{ben3}).
\begin{table}
\centering

\begin{tabular}{llcccr}
\hline \hline\noalign{\smallskip}
        & $X$
        & $\sigma_{\rm U}$
        & $\sigma_{\rm V}$
        & $\sigma_{\rm W}$
        & $V_{\rm asym}$ \\
        &
        & \multicolumn{4}{c}{---------- [km~s$^{-1}$] ----------}     \\
\noalign{\smallskip}
\hline\noalign{\smallskip}
   Thin disk   & 0.90   & $~~35$  & 20    & 16    & $-15$    \\
   Thick disk & 0.10   & $~~67$  & 38    & 35    & $-46$    \\
   Halo       & 0.0015 & $160$   & 90    & 90    & $-220$   \\
\hline
\end{tabular}{\smallskip}{\smallskip}{\smallskip}{\smallskip}\caption{
        Characteristic velocity dispersions ($\sigma_{\rm U}$, $\sigma_{\rm
        V}$, and $\sigma_{\rm W}$) in the thin disk, thick disk, and halo. $X$
        is the estimated observed fraction of stars for the given population
        in the solar neighborhood and $V_{\rm asym}$ is the asymmetric drift
        with respect to the LSR (values taken from Bensby et al. \cite{ben3}).  }

\label{tab}
\end{table}
Before applying the SFR extraction to the kinematical selected data, we have
performed the same selection also on the Nordstr\"om et al. age metallicity relation
(the AMR implemented in the SFR extraction code). The result is presented in Fig. \ref{fehk}: even if the $[Fe/H]$ dispersion slightly decreases for
lower stellar velocities, this value is still very high and no trend
is recognizable in the AMR.
For this reason, we adopted the same AMR used for the full sample without
kinematic selection. This is related to the debate over the existence of a
distinct chemical history for disk and thick disk. It is well known that high
velocity stars belong to more extended structures (thick disk and halo).
Sandage (\cite{sand}) and Casertano et al. (\cite{case}), in particular, used
kinematics to trace the thick disk population, but it is much less obvious
that these stars reveal an age-metallicity relation that can be distinguished
from that of the disk. Metallicity distributions of the thick disk and thin
disk do not allow for an unequivocal separation. Some authors (see
e.g. Gilmore et al. \cite{gilwy}, Bensby et al. \cite{ben}) argue that the
thick disk is a completely kinematically and chemically distinct Galactic
zone; e.g.  Bensby et al. (\cite{ben}) determine a specific AMR, while a
different conclusion is reached e.g. by Norris \& Green (\cite{nor89}) and
Norris \& Ryan (\cite{nor91}), who argue that the thick disk is the high
velocity dispersion tail of the old disk.

\begin{figure}
\centering
\includegraphics[width=7cm]{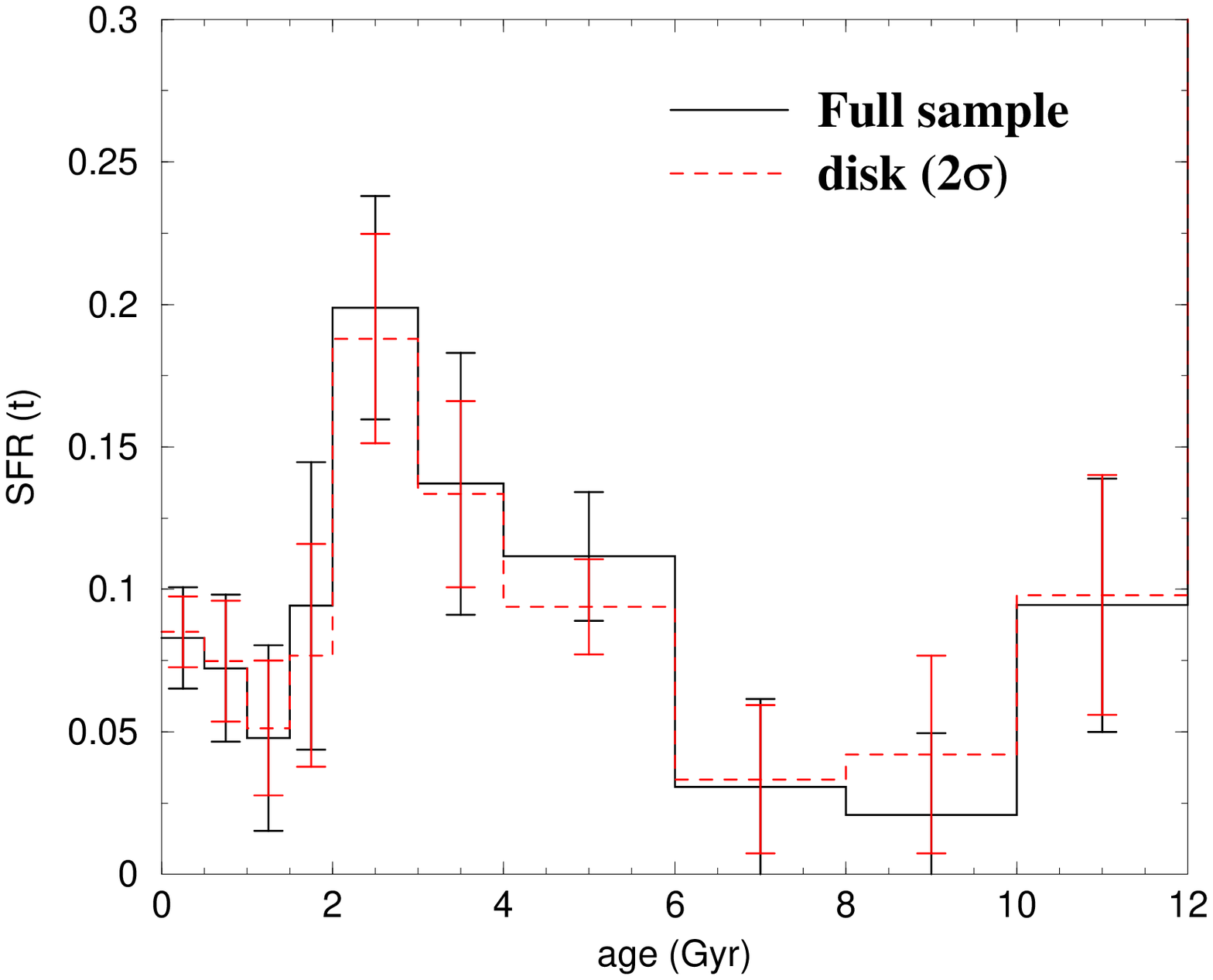}\\
\includegraphics[width=7cm]{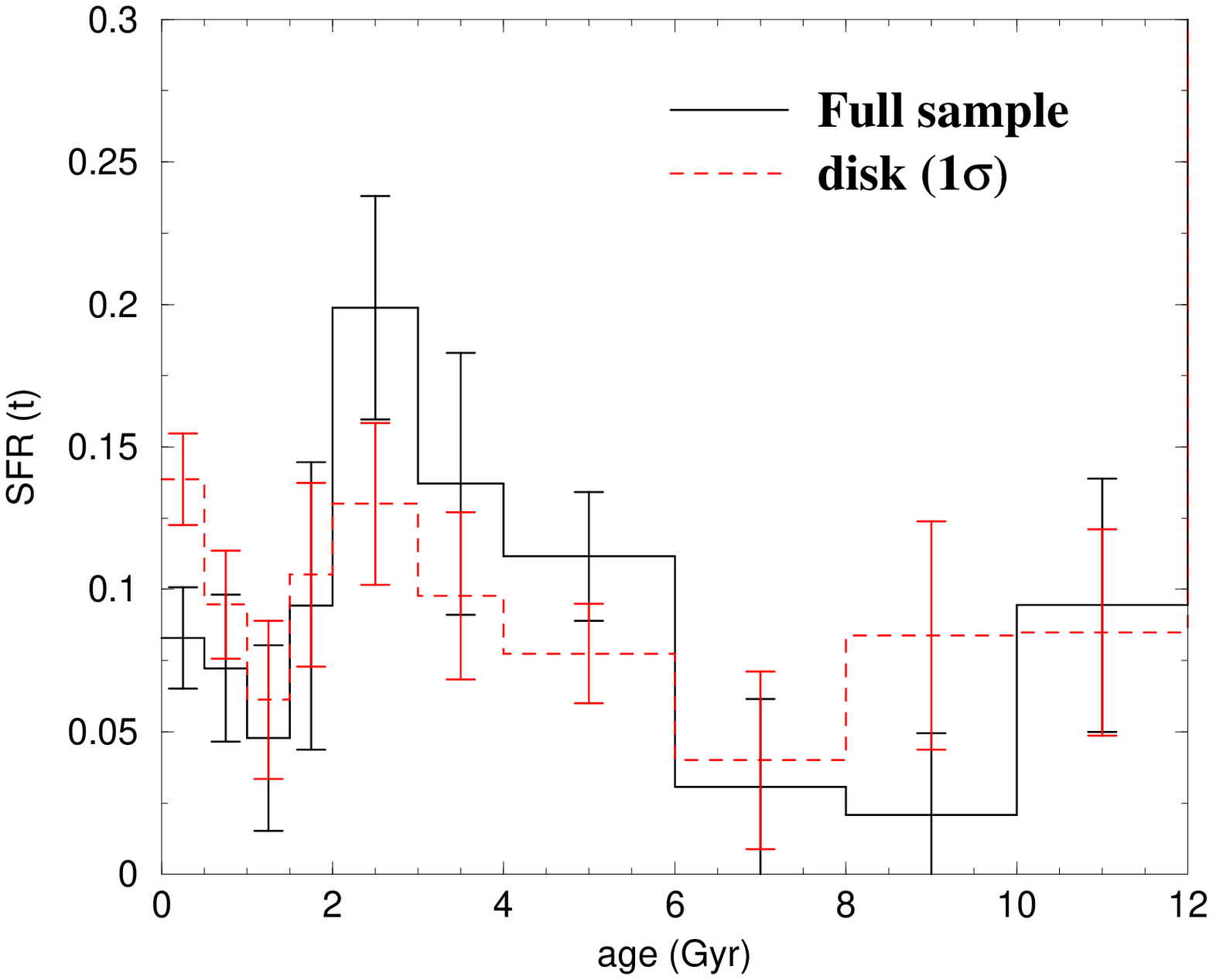}
\caption{Solid line: the recovered SFR using the full sample. Dashed line: the SFR recovered from stars with Galactic
  velocities within the velocity ellipse at $2\sigma$ (upper panel) and at
  $1\sigma$ (lower panel) the velocity ellipse of the thin disk.}
\label{lucy_sfr_kin}
\end{figure}

Figure \ref{lucy_sfr_kin} shows the recovered SFRs after 15 R-L restorations
when the sample is kinematically selected. In particular, the result in the
upper panel is found selecting objects with Galactic velocities within the velocity
ellipse at $2\sigma$ for the thin disk. Result in the lower panel is found for
objects with velocities within the velocity ellipse at $1\sigma$ for the thin
disk.

The cut at $2\sigma$ excludes essentially all halo and thick disk objects. In this
case (Fig. \ref{lucy_sfr_kin}, upper panel), the recovered SFR is almost
identical to the one without any selection. One explanation is that the
contribution of thick disk and halo stars, for the period 1-8 Gyr, is
minimal. This result confirms the general finding that the thick disk, if it
exists, seems older than thin disk. For example Fuhrman (\cite{fuh}) indicates 8 Gyr
the thick disk age while Soubiran \& Girard (\cite{soub}) from 7 to 13 Gyr (with an
average of $9.6\pm 0.3$ Gyr). In addition, the number density of local thick
disk stars is a small fraction ($\sim 8\%$) of the thin disk stars. This
result is confirmed by many works: Gilmore \& Reid (\cite{gil83}) and Chen (\cite{chen}) find
2\%, Robin et al. (\cite{rob96}) find 6\%, Soubiran et al. (\cite{soub03}) find 15\%.

In contrast, by removing stars out of $1\sigma$ one should exclude:
\begin{itemize}
\item{low velocity tails of halo and thick disk stars;}
\item{disk stars whose orbits explore large scale heights (200-300 pc);}
\end{itemize}
In this case, the recovered SFR (Fig. \ref{lucy_sfr_kin}, lower panel) has a slightly lower peak at 3 Gyr, while the activity in the last 1.5 Gyr
is increased. However, the variations are within the statistical uncertainties
(between 1 and 2 $\sigma$ of acceptance) and the global trend is preserved.
In conclusion, the recent SFR (last 6 Gyr) seems not to suffer of a
significant dynamical diffusion. In this case, a correction for a possible
disk depletion due to fast stars, does not really matter: within our level of
acceptance, the recovered SFR is a genuine local SFR and not a mere local age
distribution.

\begin{figure}
\centering
\includegraphics[width=7cm]{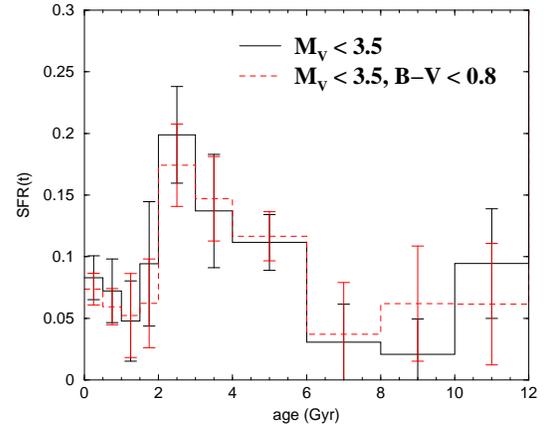}
\caption{The recovered SFR obtained from the full sample
  (solid line) and from a selection of stars with $B-V<0.8$ (dashed line).} 
\label{cnc}
\end{figure}

Because of the theoretical difficulties to reproduce the red clump stars (see
the discussion in chapter 4), the analysis was repeated excluding all stars
with $B-V>0.8$. In this case, because the excluded region involves stars of
all ages, the recovered SFR (see Fig. \ref{cnc}) is slightly different at
all ages (but still within the $1\,\sigma$ uncertainties), with a major effect
around 10 Gyr.

\section{Sensitivity to the adopted $(Z/X)_{\odot}$ value}

Recent analysis of spectroscopical data using three dimensional hydrodynamic
atmospheric models (see Asplund, Grevesse \& Sauval \cite{aspl} and references
therein) have reduced the derived abundances of CNO and other heavy elements
with respect to previous estimates (Grevesse \& Sauval \cite{greve98}, GS98).  Thus the
$Z/X$ solar value decreases from the GS98 value $(Z/X)_{\odot}=0.0230$ to
$(Z/X)_{\odot}=0.0165$. GS98 already 
improved the mixture by GN93, widely adopted in the literature ($(Z/X)_{\odot}=0.0245$),
mainly revising the CNO and Ne abundance and confirming the very good
agreement between the new photospheric and meteoric results for iron. As
already discussed our tracks are calculated for the GN93 solar mixture.

\begin{figure}
\centering
\includegraphics[width=7cm]{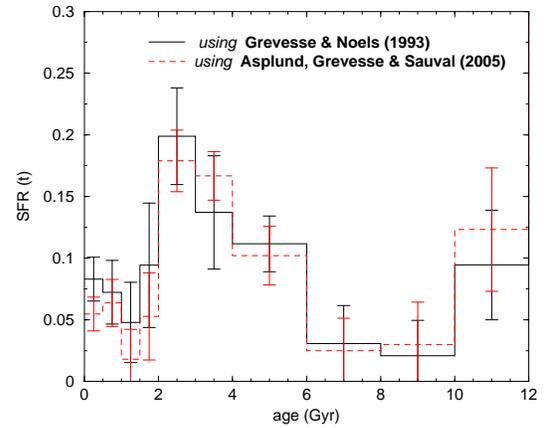}
\caption{Results for the recovered SFR obtained adopting the $Z/X$
solar value by Asplund, Grevesse \& Sauval (\cite{aspl}) (dashed line) and by GN93
(solid line).} 
\label{asgre}
\end{figure}

The change of the heavy element mixture can have two main effects: 1) the
change of theoretical tracks at fixed metallicity (but this has been shown to
be negligible; see Degl'Innocenti, Prada Moroni \& Ricci \cite{ric}); 2) the
variation of the inferred metallicity from the observed $[Fe/H]$. This could
be important for our purposes due to the adoption of the observational
age-$[Fe/H]$ relation. Figure \ref{asgre} compares the recovered SFR obtained
using the Asplund, Grevesse \& Sauval \cite{aspl} or the GN93 solar mixture; the differences are
within the statistical uncertainties.

\section{Discussion and Prospects for Further Studies}
\label{outlook}
Our recovered SFR can be now compared with other recent published
investigations. Bertelli \& Nasi (\cite{bert}), using a similar sample
(Hipparcos stars within 50 pc and brighter than $M_V=4.5$) and a similar
technique, found a local SFR that is independent by the chosen IMF, with the
exception of low exponents (the value 1.3 is rejected). The figure
\ref{confbert} shows that our derived SFR is consistent with the ones by
Bertelli \& Nasi 2001. The small discrepancies could arise from the different
inputs of the two models:
\begin{figure}
\centering
\includegraphics[width=7cm]{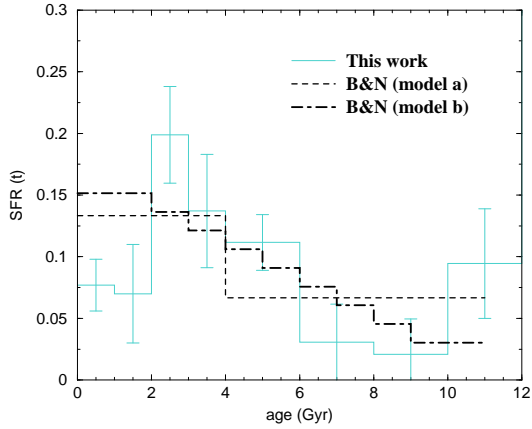}
\caption{Our recovered SFR (dotted line) rebinned for
comparison with Bertelli \& Nasi 2001 (they used two functionally different
assumptions for the star forming history: model a adopts a discontinuous
change between two constant intervals of SFR; model b adopts a linearly
increasing/decreasing SFR with a discontinuity in the slope at some time).}
\label{confbert}
\end{figure}

\begin{itemize}
\item {The adopted evolutionary tracks: Bertelli \& Nasi used the Padua
  stellar evolutionary tracks (Girardi et al. \cite{gira}) which includes
  overshooting with an efficiency of about $0.12\,H_P$ in the mass range
  $1.0\,M_{\odot}< M<1.4\,M_{\odot}$ and $\approx 0.25\,H_P$ for higher
  masses. Our code implements the Pisa stellar tracks (Cariulo et al. \cite{car},
  Castellani et al. \cite{cast03}, Castellani, Degl'Innocenti \& Marconi \cite{cast99a}). Even if
  the red clump region is poorly reproduced by the Pisa stellar tracks, while
  the Padua tracks match better, we find (see Fig. \ref{cnc}) that clump
  and giant stars have a low impact on the recovered SFR for the last 6 Gyr.}
\item{In the Bertelli \& Nasi model the stars are uniformly distributed in
the metallicity range $0.008\,<\,Z\,<\,0.03$. In contrast, we adopted the
observational AMR by Nordstr\"om et al. (\cite{nord}). Thus, their mean composition
(solar) is metal richer than ours (using Grevesse \& Noels \cite{greve},
the mean $[Fe/H]$ value $\sim -0.15$ corresponds to $Z\approx 0.012$).}

\item{Bertelli and Nasi adopt between 30
 and 70 percent of binaries (``decreasing from 70 percent  for the more
 massive primaries to about 27 percent at the faint limit $M_V$=4.5''), while our
 models are without binaries. 
In the main sequence, the luminosity of a star depends on the mass, thus a binary
 system can mimic a different mass (and a different age); we already showed
 that reasonable binary fraction has no influence on the results (i.e. section \ref{Binaries - SFR degeneracy}),
 however the amount of binaries introduced by the authors could lead to some
 differences.}

\end{itemize}

Schr\"oder \& Pagel (\cite{scho}) also used Hipparcos stars within 100 pc and
within 25 pc of the Galactic midplane. The SFR and the IMF are inferred by
comparing the expected and observed numbers of stars in particular
evolutionary phases (upper main sequence, clump, subgiants, etc.). These
authors implemented the evolutionary tracks by Eggleton (\cite{egg}) for solar
metallicity. The presence of different chemical composition was taken into
account by smearing the single metallicity CMD with a Gaussian spread.  Their
result is a local SFR that slowly increases towards recent times. The authors
explain this as an effect of a dilution of the thin disk stars as they diffuse
into larger scale heights by dynamical diffusion. In order to transform to a
column-integrated SFR they adopted a dynamic diffusion timescale of about 6
Gyr.  The final result is only slightly different from the local SFR (except
the recent 1 Gyr, where the authors correct for a radial mixing). \emph{In
practice, this result seems to confirm our finding that the dynamical
diffusion of orbits has a low impact up to 6 Gyr before the present.}

\begin{figure}
\centering
\includegraphics[width=7cm]{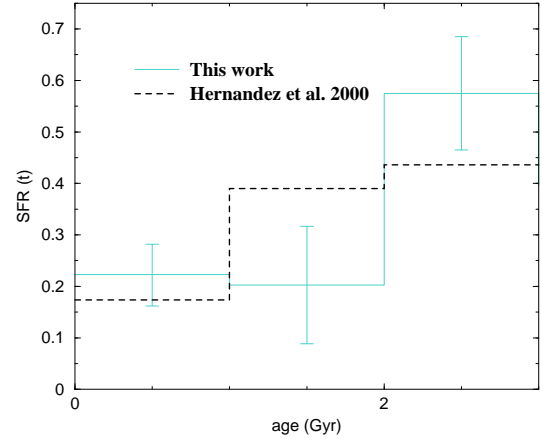}
\caption{Our recovered SFR (dotted line) compared with the Hernandez et
  al. (\cite{her2}) SFR (dashed line). Both the SFR are rebinned at 1 Gyr.} 
\label{confher}
\end{figure}

Within a Bayesian methodology, Hernandez et
al. (\cite{her2}) used an inversion method on the Hipparcos stars
brighter than $M_V=3.15$, deriving the local SFR for the last 3 Gyr. The
implemented evolutionary tracks are the Padua isochrones (Girardi et
al. \cite{gira}) with $[Fe/H]=0$. In figure \ref{confher}, it is shown our SFR
against their findings for the
last 3 Gyr (due to the lower temporal resolution of our SFR we needed to rebin
the the higher time resolution of Hernandez et al. SFR).

The two results are compatible, although our time resolution does not
allow us to resolve the SFR behavior found by these authors (a cyclic
pattern with a period of 0.5 Gyr). Considering that their sample is very
similar ours, the differences in the result could be addressed to:
\begin{itemize}

\item{Hernandez et al. (\cite{her2}) implemented a solar value ($[Fe/H]=0$) without spread,
  while we have adopted the Nordstr\"om et al. age metallicity relation;} 

\item{They used Padua isochrones (Girardi et
  al. \cite{gira}), the same as Bertelli \& Nasi (\cite{bert});}  

\item{They implemented a power law IMF with exponent 2.7 (steeper than our
  value 2.35).} 

\end{itemize}

Moreover, because their technique is very different from our method, this 
agreement constitutes an independent verification of our method.

\begin{figure}
\centering
\includegraphics[width=7cm]{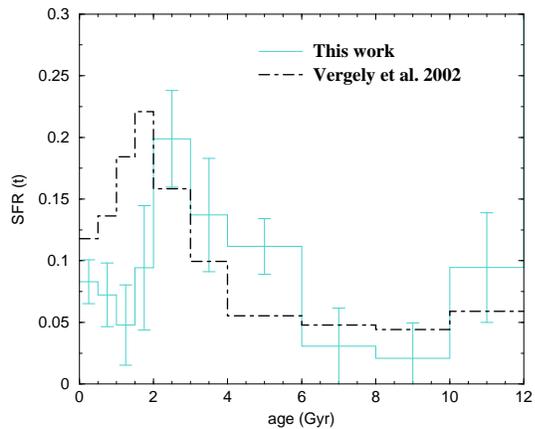}
\caption{Vergely et al. (\cite{verg}) recovered SFR (heavy dot-dashed line) compared with our
  result (dotted line). } 
\label{verga}
\end{figure}

Vergely et al. (\cite{verg}) used a similar inversion method. These authors determined
simultaneously the star formation history, the AMR and the IMF from the
Hipparcos stars brighter than $V=8$. The authors adopt a much larger sample
(not magnitude limited) and the AMR is not constrained. Their result is a
column SFR. The surprising feature is the similarity between their result (not
local), see Fig. \ref{verga}, and our SFR (local). In particular, their
column-integrated SFR decreases with lookback time on a timescale of 4-5 Gyr,
essentially the result we obtain.

This supports our result and can mean:
\begin{itemize}
\item the local stellar population is not depleted in the past, but the
  derived SFR represents a genuinely lower activity;
\item no significant dynamical diffusion has taken place on a time scale of
  4-5 Gyr.
\end{itemize}

Rocha-Pinto et al. (\cite{roch}) provided a SFR based on chromospheric emission ages
for a sample of solar-like stars within 80 pc. Their result shows
enhanced SFR episodes at 0-1 Gyr and 2-5 Gyr, that are approximately similar
to our result, and at 7-9 Gyr (but it could be a spurious effect
due to the low chromospheric emission for these ages). Also these authors find
that the effect of dynamical orbit diffusion is not severe and does not
affect the general trend of the SFR.

In conclusion, our result seems to represent a realistic SFR of the solar
neighborhood. The recovered SFR is quite independent of the kinematical
selections, suggesting that all the stellar generations (in the last 6 Gyr)
are well represented and stars are not diffused in a larger volume. The SFR is
consistent with other studies based on similar samples and different
techniques. \emph{The result that our local SFR is close to the column SFR
(Vergely et al. \cite{verg}) seems to indicate that our result is not
local and may be valid for the whole disk.}

Having checked that dynamical diffusion has not been so efficient in the
last 5-6 Gyr and the internal assumptions of the model (IMF, binaries, adopted
solar mixture) have a low impact on the result, we can discuss the physical
implication of our results.

The SFR obtained in the present work is concentrated in the recent 4 Gyr, that
is a timescale longer than Galactic disk rotation ($<1$ Gyr). This result
essentially rules out the possibility that this phenomenon is local,
suggesting Galactic scale triggering event. It is difficult to explain our
result if the Galactic disk is a ``closed box'' (see e.g. Van den Bergh
\cite{van}, Schmidt \cite{schm}): in this case the resulting SFR would be
\emph{decreasing} from the disk formation to the present (in opposition to our
result), following the normal exhaustion of the gas content and an increased
production of inert remnants. Even if the disk is periodically refilled with
gas, our result is difficult to explain: the resulting SFR would be nearly \emph{constant} in time (unless the infall is huge, but in this case
the age-metallicity relation would change relative to observational evidence;
e.g. Valle et al. \cite{valle}).

Thus, the recovered SFR seems to indicate some kind of \emph{induced event},
for example by the accretion of a satellite galaxy. However, the tracks left
by a such an intruder should be recognized in the age-metallicity relation,
while the survey of Nordstr\"om et al. (\cite{nord}) shows practically no change in
mean metallicity from 1 to 12 Gyr. An accretion should be evident
from the analysis of the kinematical properties of stars in different age bins, but the
methods to obtain stellar ages are still affected by large errors (see
discussion in section \ref{nordmeth}).

Much larger surveys of stellar ages and metallicities as a function of
galactocentric distance and kinematics are needed to test our hypothesis:
comparing results from different regions of the disk could make clear if the
recovered event is really a global event.

\section{Conclusions}

We have used a selection of the Hipparcos stars to recover the local SFR. The
analysis is restricted to the stars within 80 pc and brighter than
$M_V=3.5$. Numerical experiments with artificial CMDs show that, at these
luminosities, neither the IMF nor the binary fraction are critical inputs,
while the possibility to recover the SFR is strongly influenced by the adopted
age-metallicity relation.

In particular, this result was checked assuming the observational AMR for the
solar neighborhood by Nordstr\"om et al. (\cite{nord}): the simulation with
artificial CMDs indicate that most of the information about the underlying SFR
is still recoverable.  Finally, we applied the algorithm to real Hipparcos
data. In contrast with artificial CMDs, the first problem was the presence of
observational uncertainties (due to photometric and parallax errors).  To take
these uncertainties into account, we applied to the data the Richardson-Lucy
technique as introduced in Cignoni \& Shore (\cite{cign}), cleaning the Hipparcos CMD from the
observational errors.  Then, assuming the observational AMR by Nordstr\"om et
al. (\cite{nord}), we have found the most probable SFR from our sample. The result
indicates that \emph{the recent local SF history of the Galactic disk is
increasing from the past to the present with some irregularities}. The mean
value increases very steeply from 6-7 Gyr ago up to 2 Gyr, in a way
qualitatively similar to the findings of Vergely et al. (\cite{verg}) and Bertelli
\& Nasi (\cite{bert}). In particular, this result is is quite independent against
kinematic selections, suggesting that:
\begin{enumerate}
\item The local contamination of
halo and thick disk stars is negligible in the last 6 Gyr and/or
these populations are older than 6 Gyr; 
\item In the last 5-6
Gyr, all the stellar generations are well sampled; in other
words, the recovered local SFR seems not to be biased by dynamical diffusion and the
local volume is not ``depleted'' by old disk stars. 
\end{enumerate}
The timescale of the recovered SFR seems too long (larger than the
  dynamical timescale) to be attributed to local events: an accretion of a
  satellite galaxy is suspected.

\begin{acknowledgements}
We warmly thank C. Chiosi and J. K\"oppen for their very useful suggestions
regarding the PhD thesis by M. Cignoni, and M. Bertero, G. Bono, P. Franco,
M. Martos, and G. Valle for discussions. Financial support for this work was
provided by the National Institute of Astrophysics (INAF). We dedicate this paper to the memory of Prof. Vittorio Castellani.
\end{acknowledgements}

\end{document}